\documentclass[%
reprint,
 amsmath,amssymb,
 aps,
pre,
floatfix
]{revtex4-1}

\usepackage[
margin=1in,
]{geometry}

\newcommand{\mytitle}{{Analytic results and weighted Monte Carlo simulations for CDO pricing}}

\usepackage{indentfirst}%

\usepackage[sort&compress]{natbib}

\usepackage[english]{babel}
\usepackage[utf8x]{inputenc}

\usepackage{amsmath}%
\usepackage{amsfonts}%
\usepackage{amssymb}%
\usepackage{enumerate}%

\usepackage{watermark}

\usepackage[usenames,dvipsnames]{color}%
\usepackage{ifpdf}%
\ifpdf
    \usepackage[pdftex]{graphicx}
    \usepackage[pdftex,unicode]{hyperref}
    \usepackage[all]{hypcap}
    \hypersetup{
        bookmarks=true,         
        unicode=true,          
        pdftoolbar=true,        
        pdfmenubar=true,        
        pdffitwindow=true,      
        pdfstartview={FitH},    
        pdftitle={\mytitle},    
        pdfkeywords={CDO},
        pdfnewwindow=true,      
        colorlinks=true,        
        linkcolor=BlueViolet,   
        citecolor=OliveGreen,   
        urlcolor=blue           
    }
\else
    \usepackage[dvips]{graphicx}
    \usepackage[dvips,unicode]{hyperref}
    \providecommand{\hyperref}[2][]{#1 \ref{#2}}
    \providecommand{\pdfbookmark}[3][Ignorer]{}
    \providecommand{\texorpdfstring}[2]{#1}
\fi

\def\br #1{\left( #1 \right)}
\def\Br #1{\left[ #1 \right]}
\def\BR #1{\left\{ #1 \right\}}

\newcommand{\1}{\ensuremath{1\!\!1}}
\renewcommand{\P}{\mathbf{P}}
\newcommand{\E}{\mathbf{E}}
\newcommand{\D}{\mathbf{D}}
\newcommand{\var}{{\mathbf{Var}}}

\providecommand{\ud}{\mathrm{d}}
\renewcommand{\d}{\ud} 
\providecommand{\min}{{\mathrm{min}}}
\providecommand{\max}{{\mathrm{max}}}

\newcommand{\Rel}{}
\newcommand{\Alt}{{'}}

\newcommand{\ON}{{ON}}
\newcommand{\RN}{{Ra\-don\,--\,Ni\-ko\-d\'ym }}
\newcommand{\ha}{{\mathfrak{a}}}
\newcommand{\hd}{{\mathfrak{d}}}
\newcommand{\defPV}{{\texttt{defPV}}}

\newcommand{\premPVbp}{{\texttt{premPV1bp}}}
\newcommand{\ind}{{\1}} 
\newcommand{\I}{{\1}} 
\newcommand{\Xdef}{X^{({\tt def})}}
\newcommand{\Xprem}{X^{({\tt prem})}}
\newcommand{\logar}{\ln}
\newcommand{\coloronline}{(color online)}
\newcommand{\Rec}{\ensuremath{\tilde r}}

\begin{document}

\title{\mytitle}
\author{Marcell Stippinger}
\affiliation{Dept.\ of Theor.\ Phys., Budapest Univ.\ of Tech.\ and Econ.}
\author{B\'alint Vet\H{o}}
\affiliation{Dept.\ of Stochastics, Budapest Univ.\ of Tech.\ and Econ.}
\author{\'Eva R\'acz}
\affiliation{Dept.\ of Theor.\ Phys., Budapest Univ.\ of Tech.\ and Econ.}
\author{Zsolt Bihary}
\affiliation{Morgan Stanley Hungary}

\date{\today}

\begin{abstract}
We explore the possibilities of importance sampling in the Monte Carlo pricing of a structured credit derivative referred to as \emph{Collateralized Debt Obligation} (CDO).  
Modeling a CDO contract is challenging, since it depends on a pool of (typically $\sim 100$) assets, Monte Carlo simulations are often the only feasible approach to pricing. Variance reduction techniques are therefore of great importance. 
This paper presents an exact analytic solution using Laplace-transform and MC importance sampling results for an easily tractable intensity-based model of the CDO, namely the compound Poissonian. Furthermore analytic formulae are derived for the reweighting efficiency. The computational gain is appealing, nevertheless, even in this basic scheme, a phase transition can be found, rendering some parameter regimes out of reach. A model-independent transform approach is also presented for CDO pricing. 
\end{abstract}

\maketitle
\section{Introduction}\label{sec:Introduction}
Econophysics literature, especially due to the availability of high-resolution stock exchange trading data, has initially been concerned with interpreting equity stylized facts \cite{Mantegna2000, Cont2001, Kertesz2011a} and equity derivatives.
The past decade however, has shown a tremendous rise in the trading volume of credit derivatives \cite{sifma}, i.e., products depending on an event like bankruptcy, default or changes in the credit rating of a company or government. The buyer of the protection against such an event transfers his credit risk to the seller, and pays a periodic fee in return, maximally until the maturity of the contract. Although in this setting, credit derivatives are instruments of risk reduction, since it is not necessary to own, e.g., a bond of the companies of interest, they open ground for speculation, too. 

The simplest credit derivative is the \emph{Credit Default Swap} (CDS), which is a swap transferring the risk of holding a fixed income product of a single company, such as a bond. In case the company defaults on paying the bond coupons, the buyer of the CDS is entitled to the face value of the bond. The price (the periodic payment to the seller) of a CDS is referred to as \emph{CDS spread}. The higher the spread, the riskier the market deems investing in the company in question.

The simplest credit derivative is the \emph{Credit Default Swap} (CDS), which is a swap transferring the risk of holding a fixed income product of a single company, such as a bond. In case the company defaults on paying the bond coupons, the buyer of the CDS is entitled to the face value of the bond. The price (the periodic payment to the seller) of a CDS is quoted in bps, i.e., $10^{-4}$ of the nominal value of the contract, and is referred to as \emph{CDS spread}. The higher the spread, the riskier the market deems investing in the company in question. Opposed to CDS-s, structured products depend on the status of many underlying assets (e.g., the bonds of many companies) which, due to the interwoven nature of business relationships and macroeconomic factors, have a complex correlation structure. The subprime mortgage crisis of 2007--2008, has shown that the rising volume of such contracts \cite{sifma} can lead to unforeseen instabilities. 

The complexity of the stochastic models used for pricing structured products are often not analytically solvable,  Monte Carlo simulations are essential tools for quantitative analysts.
Much effort has been spent on improving different aspects of these MC simulations, like speed and accuracy. An additional important task is calibrating the model parameters to market-observable prices of benchmark instruments. One approach to the latter problem is reweighting MC paths gained using a prior probability measure (weighted Monte Carlo, or WMC, Avellaneda \emph{et al.}\ \cite{Avellaneda2001a} and in the context of credit derivatives Cont \emph{et al.}\ \cite{Cont2009a}).
Our work also involves reweighting MC paths, but the goal is reducing the variance of the Monte Carlo estimates of the expected cashflow, not calibration to market observables. The reweighting scheme is based on the \RN derivative, and is also referred to as importance sampling (see Sec.\ 4.3 in \cite{Staum2003a}).

In this paper, we are dealing with collateralized debt obligations,
which are contingent on the default status of the constituents of a reference
portfolio, such as Markit iTraxx Europe or CDX NA IG\cite{markit}. The contract can basically be viewed as a combination of many CDS-s, however, the net loss on the portfolio is cut up into smaller intervals termed \emph{tranches}. The seller of a CDO
tranche pays the excess loss on the portfolio above a threshold (attachment
point of the tranche), up to a maximum value (detachment point), and receives in return a
periodic payment from the buyer (proportional to the remaining width of the tranche),
referred to as CDO tranche \emph{premium}. Standard CDO tranches for the CDX NA IG series include the \emph{equity} ($0-3\%$), the \emph{mezzanine} ($3-7\%$, $7-15\%$) and the \emph{super senior} ($15-100\%$) tranches. The attachment points of the super senior tranches of other indices range from $15\%$ to $35\%$; in our work, we used the $30-100\%$ slice as a representative super senior tranche. In the market, these products are  quoted in either basis points ($10^{-4}$, e.g., a spread of 100 basis points means the annual premium payment fraction, although payments are typically made semi-annually), i.e., the value of the periodic payment, or, assuming a fixed premium, the value of the upfront payment (in \% of the tranche notional). For the sake of simplicity, in this paper we assume zero upfront in each case considered.

Note that the cost functions of buyer/seller are
non-linear, thus, the dependence structure between portfolio elements plays a
crucial role in pricing a CDO tranche. ``Bottom-up'' approaches, including
the Gaussian copula model \cite{Li2000} which became infamous during the recent
financial crisis \cite{Salmon2009}, try to estimate this dependence structure,
and price CDO's consistently with single name credit defaults swaps (CDS's,
these depend on the default status of a single company). ``Top-down'' models,
in contrast, deal directly with the aggregate loss on the portfolio, thereby
decreasing the number of model parameters (in bottom-up approaches, this is
done by introducing homogeneity assumptions) and giving up information about
component risks (but see the ``random thinning'' procedure in
\cite{Giesecke2009a}). In this paper, we consider a simple top-down compound 
Poisson model in order to retain analytic tractability and demonstrate 
MC possibilities.

The paper is organized as follows: Sections \ref{par:CDO} and
\ref{sec:quantities} summarize model details and the relevant quantities.
Section \ref{sec:AnalyticApproach} introduces a general method for CDO pricing
for models including a constant interest rate and deduces analytic formulas for
the cash flow of the CDO contract in the Poissonian case. Section \ref{sec:MonteCarloResults} turns to presenting the possibilities of the Monte Carlo simulation, and presents the path-reweighting technique, which, for this simple model, can again be analytically verified.

\section{Basic concepts}\label{sec:concepts}
\subsection{Collateralized Debt Obligation -- The basic model} \label{par:CDO}

Let us assume that the CDO is based on a pool of $N_{\text{comp}}$ companies, which, for the
sake of simplicity, corresponds to a total notional of $N \equiv 1\,\text{USD}$.

This model is based on a single default process $D_t$ which is a compound
Poisson process in the following sense: the default events occur according to a
simple Poisson process of intensity $\rho$, and during the $i^{\text{th}}$ event, a fraction $J_i$ of the
companies default. The jump sizes $J_i$ are independent and identically
distributed random variables of exponential distribution with parameter $\lambda$,
i.e., $\P\br{J_i < x} = 1 - e^{- \lambda x}$. We also use the notation
$\mu=\lambda^{-1}$ for the expected value of the jumps. We assume that $\BR{J_i}_i$ are also
independent from the jump times.

We considered two natural ways to define the actual loss process with values in
$[0,1]$. The first one is referred to as the linear specification given by
\[L_t^\textup{lin}:=\min\left(D_t, 1\right).\]
The exponential specification
\begin{equation}
L_t^{\exp} := 1 - \exp\BR{-D_t}\label{Ldef}
\end{equation}
is obtained by a smooth transformation from $D_t$. For small values of $D_t$,
the two quantities $L_t^{\exp}$ and $D_t$ are close, which is the typical case
for the relevant parameter regimes.

In what follows, we use the exponential specification (unless otherwise
indicated) and denote the loss $L_t^{\exp}$ by $L_t$ for simplicity, and we say
that, at time $t$, an $L_t$ proportion of the companies defaulted. (Note that
in each specification, the portfolio loss is continuous. This is a natural simplification for typical index portfolios where $N_{\text{comp}}$ is $\sim 100$.)

The buyer of the a CDO tranche $\Br{a, d}$ makes periodic payments (called \emph{premium leg}) proportional
to the \emph{outstanding notional} (the remaining width) on the tranche until either
the maturity $M$ (the expiry of the contract) is reached or the loss exceeds the detachment point.
The seller of the protection pays the \emph{default leg} after each
default event, which is the increment of $\min\{L_t, d\}-\min\{L_t,a\}$.

The default of a company does not mean that it becomes entirely worthless,
a fraction $\Rec$ of its original value is recovered,
i.e., the portfolio loss $L_t$ increases by a $1-\Rec$ proportion of the jump
which occurred at time $t$. In the simplest setting, the recovery $\Rec$ is a
deterministic constant value $\Rec\in[0,1)$ (see \cite{Li2000, Burtschell2005a,Burtschell2009a} for example).
In this paper, we simply assume that $\Rec = 0$.

\subsection{Relevant quantities}\label{sec:quantities}

Let the interest rate $r$ be constant in time. For a tranche $[a,d]$, we denote by
\begin{equation}
\ell_t^{a,d}= \min\br{L_t, d} - \min\br{L_t, a}\label{eq:elldef}
\end{equation}
the loss on this tranche at time $t$. The phrase tranche loss is often
used in the literature for $\ell_M^{a,d}$ the total loss at the maturity.

The default leg present value ({\tt defPV}) of a tranche is approximately the expected present
value of the tranche loss, more precisely, the increments of the loss are discounted.
In mathematical terms,
\begin{equation}
{\tt defPV}=\E\left(\int_0^M e^{-rt}\,\d\ell_t^{a,d}\right)\label{defpvdef}
\end{equation}
which is meant as a Stieltjes integral. The dependence of the {\tt defPV} on the
tranche is suppressed in the notation.

The premium leg present value ({\tt premPV}) is the expected present value of the
total amount of the periodic payment by the protection buyer. The annual
payment is ${\tt spread}\times ON_t$ where the {\tt spread} is given in basis points ({\tt
bp}s) and fixed in the contract. $ON_t$ is the outstanding notional of the
tranche $[a,d]$ at time $t$, i.e.
\[ON_t=d-a-\ell_t^{a,d}.\]
Hence,
\begin{equation}
{\tt premPV}={\tt spread}\times\E\left(\int_0^M e^{-rt} ON_t\,\d
t\right)\label{prempvdef}
\end{equation}
where the expectation on the right-hand side is denoted by {\tt premPV1bp}.

The aim of CDO pricing is to give a good estimate of the fair value of the {\tt
spread} (and {\tt upfront}) for given tranches. Therefore, the equation
\[\texttt{defPV}=\texttt{spread}\times\texttt{premPV1bp}\,\,(+\texttt{upfront})\]
has to be satisfied, since the left-hand side is the expected income of the
protection buyer, whereas the right-hand side is that of the protection seller.
The problem is now reduced to finding the values {\tt defPV} and {\tt
premPV1bp}.

Throughout the paper, we will also use the notation
\begin{align}
\Xdef:&=\int_0^M e^{-rt}\,\d\ell_t^{a,d},\label{eq:defXdef}\\
\Xprem:&=\int_0^M e^{-rt} ON_t\,\d t,\label{eq:defXprem}
\end{align}
which stand for the present value of the tranche loss and that of the total
amount of premium leg paid by the protection buyer respectively. Note that
these are random variables, and their expectations are
\begin{align}
\E\left(\Xdef\right)&=\defPV,\\
\E\left(\Xprem\right)&=\premPVbp,
\end{align}
compare with \eqref{defpvdef} and \eqref{prempvdef}.

We define
\begin{align}
\ha:&=-\ln(1-a),\label{eq:defha}\\
\hd:&=-\ln(1-d).\label{eq:defhd}
\end{align}
Due to \eqref{Ldef}, the event $\{L_t\text{ exceeds }a\}$ is equal to
$\{D_t\text{ exceeds }\ha\}$. The same hold with $d$ and $\hd$ respectively.
This notation serves to reduce the length of subsequent formulas.



\section{Analytic approach}\label{sec:AnalyticApproach}
In the compound Poissonian case, we can derive explicit for\-mu\-las for the relevant quantities {\tt defPV} and {\tt premPV1bp}.
The expressions contain an infinite series representation, which converges faster than exponential, therefore, our method provides a promising approach to CDO pricing.
The basic idea of the computations is that we decompose the underlying expectation according to the first passage of certain levels of loss.

\subsection{The \texorpdfstring{{\tt defPV}}{defPV} and
\texorpdfstring{{\tt premPV1bp}}{premPV1bp} expressed with the first passage time}\label{sec:FirstPassageTime}

The loss $L_t$ is an exponential transformation of the compound Poisson process $D_t$, see \eqref{Ldef}, and the computations can be done in terms of $D_t$. Hence, we introduce
\[T_h:=\min\{t\ge0:D_t\ge h\}\]
the first passage time of level $h$ for $D_t$. It turns out that the quantities \texttt{defPV} and \texttt{premPV1bp} can be expressed by the integral of the function
\begin{equation}
\varphi_r(h,M):=\E\left( e^{-rT_h} \ind(T_h < M)\right)\label{deffi}
\end{equation}
which is a modified Laplace transform of the first passage time, and
$\ind\br{\cdot}$ denotes the indicator function. Therefore, knowing the
function $\varphi_r$ is enough to determine {\tt defPV} and {\tt premPV1bp} and
also the fair value of {\tt spread} and {\tt upfront}.

\begin{figure}[htbp]
\centering
\includegraphics[width=0.95\columnwidth]{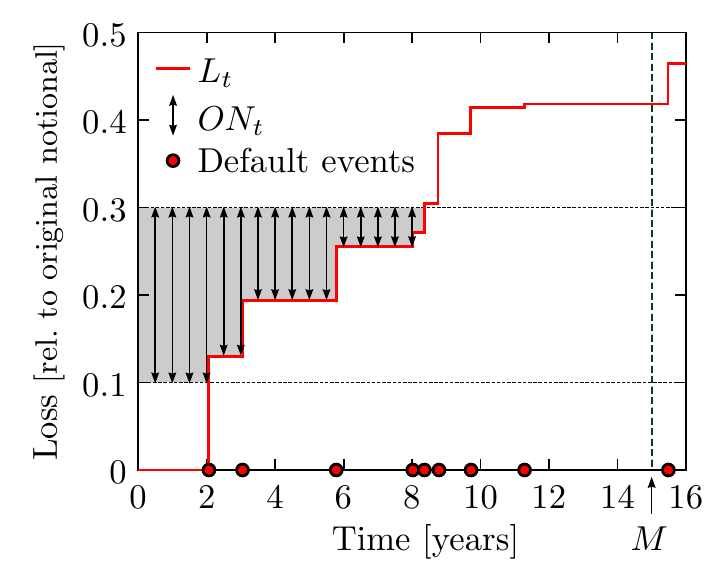}
\caption{A sample scenario of the loss process $L_t$. The jumps in the solid red line indicate the default events occurring according to a Poisson process of intensity $\rho$. Assuming continuous payment, the default leg present value {\tt defPV} is the sum of the discounted increments of the loss process $L_t$ (see Eq.\ \eqref{defpvnewdef}) and the the premium leg present value {\tt premPV1bp} is related to the grey area (in case $r = 0$, they coincide, see Eq.\ \eqref{prempvnewdef}). \coloronline}
  \label{fig:def+prem}
\end{figure}

It is not difficult to show that instead of the original definition of {\tt defPV} in \eqref{defpvdef} where the increments of loss are added, with a new approach, an integration along the vertical axis can be done, see also \hyperref[fig:def+prem]{Figure \ref*{fig:def+prem}}. We obtain
\begin{equation}\begin{aligned}
{\tt defPV}&=\E\left(\int_a^d e^{-r\min\{t\ge0:L_t\ge x\}}\I(x<L_M)\,\d x\right)\\
&=\E\left(\int_{\ha}^{\hd} e^{-rT_h} \I(T_h < M)\,e^{-h}\,\d h\right)\\
&=\int_{\ha}^{\hd} \varphi_r(h,M)\,e^{-h}\,\d h
\end{aligned}\label{defpvnewdef}\end{equation}
after a change of variable under the integral sign ($x\to h=-\logar(1-x)$),
which corresponds to considering the compound Poisson process $D_t$ itself
instead of $L_t$.

Similarly, definition \eqref{prempvdef} of the {\tt premPV1bp} is replaced by
\begin{equation}\begin{aligned}
&{\tt premPV1bp} = \\
&=\E\left(\int_a^d\int_0^{\min\{0\le t\le M:L_t\ge x\}}
e^{-rs}\,\d s\,\d x\right)\\
&=\E\left(\int_{\ha}^{\hd}\int_0^{\min(T_h,M)} e^{-rs}\,\d
s\;e^{-h}\,\d h\right)\\
&=\E\left(\int_{\ha}^{\hd}\frac{1-e^{-r\min(T_h,M)}}r e^{-h}\,\d
h\right) .
\end{aligned}\label{prempvnewdef}\end{equation}
In terms of \hyperref[fig:def+prem]{Figure \ref*{fig:def+prem}}, formulas \eqref{prempvdef} and \eqref{prempvnewdef} give the indicated area in two different ways. After straightforward manipulations,
one obtains from \eqref{prempvnewdef} that
\begin{equation}\begin{aligned}
{\tt premPV1bp}&=\frac{1}{r} \left(1-e^{-rM}\right)(d-a)\\
&\quad+\frac{e^{-rM}}{r}
\int_{\ha}^{\hd} \varphi_0(h,M)\,e^{-h}\,\d h\\
&\quad-\frac{1}{r} \int_{\ha}^{\hd} \varphi_r(h,M)\,e^{-h} \,\d h ,
\end{aligned} \label{prempvrewrite}\end{equation}
with the unknown function $\varphi_r$ depending on the details of the model. An
important remark is that formulas \eqref{defpvnewdef} and \eqref{prempvrewrite}
are valid for \emph{any} distribution of the process $D_t$, not only for the
compound Poissonian case, we assumed only a constant interest rate and
continuous payment possibilities for both sides (it is simple to generalize the
results for deterministic interest rate functions $e^{-rt} \rightarrow
\exp\{-\int_0^t r\br{s} \d s\}$, but we omit this possibility in the present
paper). Using the present approach, for any distribution of $D_t$, it is enough
to determine the function $\varphi_r$ for pricing a CDO. In the next
subsection, we calculate $\varphi_r$ in the compound Poissonian case.

\subsection{Partial differential equation for the Laplace transform of the first passage time}

For the compound Poisson process $D_t$, a series representation of $\varphi_r$ can be given as follows. In order to avoid later confusions, we fix the value of the interest rate $r$, and we suppress the subindex of $\varphi$. The expectation in \eqref{deffi} can be decomposed according to the time and the size of the first jump of the process $D_t$ since these are independent exponential random variables with parameter $\rho$ and $\lambda$ respectively. After change of variable, one can obtain the integral equation
\begin{equation}\begin{aligned}
&\varphi(h,M)\\
&\quad=\rho e^{-(\rho+r)M} e^{-\lambda h}\\
&\qquad\times\int_0^M e^{(\rho+r)y}\left(\lambda \int_0^h e^{\lambda x}
\varphi(x,y)\,\d x+1\right)\d y
\end{aligned}\label{firewrite}\end{equation}
using the memoryless property of the exponentials. Differentiating \eqref{firewrite}, we can deduce the partial differential equation
\begin{equation}
\partial^2_{hM}\varphi+\lambda\partial_M\varphi+(\rho+r)\partial_h\varphi+\lambda
r\varphi=0.\label{pde}
\end{equation}

One boundary value is obviously
\begin{equation}
\varphi(h,0)=0.\label{M=0}
\end{equation}
One has to be more careful at the other one. By definition, the function is constant $1$ along the line $h=0$, but the bivariate function $\varphi(h,M)$ is not continuous here. Therefore, we redefine $\varphi$ at the boundary, or, more precisely, we can say that the definition \eqref{deffi} is valid only if $h>0$, and we extend the function continuously. Since as $h\downarrow 0$, the probability that the first jump exceeds $h$ tends to 1, the boundary value is
\begin{align}
\nonumber \varphi(0,M) &=\lim_{h\downarrow0}\varphi(h,M)= \int_0^M e^{-ry}\rho e^{-\rho y} \d y \\
& = \frac{\rho}{\rho+r}
\left(1-e^{-(\rho+r)M}\right).\label{h=0}
\end{align}

\subsection{Solution of the PDE}

The equation \eqref{pde} is a second order hyperbolic partial differential equation, which contains extra terms of lower order. One way of solving it is performing Laplace transformation in both variables. Let
\[\varphi_{st}:=\int_0^\infty \int_0^\infty e^{-sh} e^{-tM} \varphi(h,M)\,\d h\,\d M\]
be the Laplace transform. We will also use the functions
\begin{align*}
\varphi_s(M):&=\int_0^\infty e^{-sh} \varphi(h,M)\,\d h,\\
\varphi_t(h):&=\int_0^\infty e^{-tM} \varphi(h,M)\,\d M
\end{align*}
for computing the Laplace transforms of $\partial^2_{hM}\varphi$, $\partial_h\varphi$ and $\partial_M\varphi$. They can be given by integration by parts. In the calculation, the Laplace transforms of the boundary values \eqref{M=0} and \eqref{h=0} also appear.

The Laplace transform of the equation \eqref{pde} is written as
\begin{align*}
&st\varphi_{st}-t\frac\rho{t(t+\rho+r)}+\lambda t\varphi_{st}\\
&\quad+(\rho+r) \left(s\varphi_{st} -\frac\rho{t(t+\rho+r)}\right)+\lambda
r\varphi_{st}=0.
\end{align*}
The unknown function $\varphi_{st}$ can be expressed easily:
\begin{equation}
\varphi_{st}=\frac\rho t\,\frac1{st+\lambda t+(\rho+r)s+\lambda r}.
\end{equation}
The elimination of variable $t$ can be done by using the identity
\[\int_0^\infty e^{-px}\frac{1-e^{-\alpha x}}\alpha\,\d x=\frac1{p(p+\alpha)}\]
with $\alpha=((\rho+r)s+\lambda r)/(s+\lambda)$. We get
\begin{equation}\begin{aligned}
\varphi_s(M)&=\frac\rho{\rho+r}\frac1{s+\frac{\lambda r}{\rho+r}}\\
&\qquad\times\left(1-\exp\left(-(\rho+r)M+\frac{\lambda\rho
M}{s+\lambda}\right)\right).
\end{aligned}\label{fisM}\end{equation}
The second inversion is not completely obvious. The difficulty is that, in the second term in \eqref{fisM}, the variable $s$ appears in two different places: in the denominator of the prefactor $1/(s+\lambda r/(\rho+r))$ and in the exponential as well.

We could use the general identity
\begin{equation}\begin{aligned}
&\int_0^\infty e^{-px}\left(e^{\beta x}\int_0^x f(y)\,\d y\right)\d x\\
&\qquad=\frac{\int_0^\infty e^{-(p-\beta)x}f(x)\,\d x}{p-\beta}
\end{aligned}\label{generalidLapl}\end{equation}
with $\beta=-\lambda r/(\rho+r)$ to proceed. Then the problem reduces to finding the inverse Laplace transform of the function \[g_s(M)=\exp\left(\frac{\lambda\rho M}{s+\frac{\lambda\rho}{\rho+r}}\right)\] in the $s$ variable where $s$ occurs only once. It can be solved by considering the series expansion of the exponential and by performing the inversion for each term individually using the general formula
\[\int_0^\infty e^{-px}\left(\frac{x^{n-1}}{(n-1)!}\,e^{-\alpha x}\right)\d x=\frac1{(p+\alpha)^n}.\]
One may notice that for each term of the sum in the series expansion, we get functions of the form $x\mapsto cx^{n-1}e^{-\nu x}$. These are to be integrated by the left-hand side of \eqref{generalidLapl} in place of $f$. Hence, we also use the following series representation of the lower incomplete gamma function:
\[\int_0^x t^{n-1}e^{-\nu t}\,\d t=\frac{(n-1)!}{\nu^n}\left(1-e^{-\nu x}\sum_{k=0}^{n-1} \frac{(\nu x)^k}{k!}\right).\]
The resulting formula is
\begin{equation}\begin{aligned}
\varphi_r(h,M)&=\frac\rho{\rho+r}\cdot e^{-\lambda h-(\rho+r)M}\\
&\;\times\sum_{n=1}^\infty \frac{(\rho+r)^n M^n}{n!}
\sum_{k=0}^{n-1}\left(\frac{\lambda\rho h}{\rho+r}\right)^k\cdot\frac1{k!}.
\end{aligned}\label{result}\end{equation}
It is not hard to see that the solution \eqref{result} indeed satisfies the
equation \eqref{pde} along with the boundary values \eqref{h=0} and
\eqref{M=0}. One more important special case is if $r=0$. The \eqref{result}
reduces to
\begin{equation}
\varphi_0(h,M)=\sum_{n=1}^\infty e^{-\rho M}\frac{(\rho M)^n}{n!}
\sum_{k=0}^{n-1} e^{-\lambda h}\frac{(\lambda h)^k}{k!}\label{eq:series_r=0}
\end{equation}
which can be verified intuitively as follows. The left-hand side of
\eqref{eq:series_r=0} is equal to $\P(T_h<M)$ by definition. The right-hand
side is the sum of the weights of those trajectories of $D_t$ which give rise
to the event $\{T_h<M\}$. Assume that $D_t$ has $n$ jumps in the interval
$[0,M]$. The jump sizes can be generated by a Poisson point process with
intensity $\lambda$ along the vertical axis. $T_h<M$ is satisfied if and only
if this Poisson process has $k$ point in $[0,h]$ with $0\le k<n$. In the
general $r>0$ case, the same factors as on the right-hand side of
\eqref{eq:series_r=0} appear in the formula, but there is no explicit
probabilistic interpretation of \eqref{result}.

The benefit of the computations is that, with \eqref{result}, the value of \texttt{defPV} and \texttt{premPV} are known explicitly using \eqref{defpvnewdef} and \eqref{prempvrewrite}. The expression \eqref{result} is computationally stable because of the factorial in the denominator. This new method gives an unbiased answer also for those tranches for which one cannot guarantee enough sample paths with the usual Monte Carlo simulation. In these parameter regimes (e.g., pricing a super senior tranche), the results can be compared to those of the weighted Monte Carlo simulation.

\section{Monte Carlo Simulation}\label{sec:MonteCarloResults}

\begin{figure}[htbp]
  \includegraphics[width=\columnwidth]{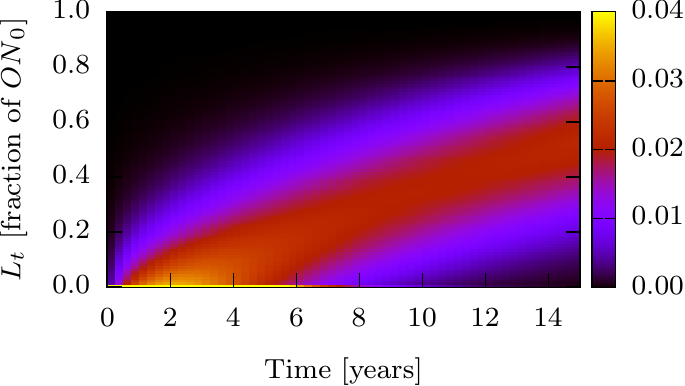}
  \caption[Loss surface example for the compound Poisson process]{Loss surface of the compound Poisson process with intensity $\rho=0.5$ and exponential jump sizes of parameter $\lambda=10$ (expected value $\frac{1}{\lambda}=0.1$) generated by $10^6$ Monte Carlo paths.
  The $L_t=0$ stripe is not shown, because it contains much larger probabilities than the central region, and would therefore obstruct visibility.
  \coloronline}
  \label{fig:losssurf}
\end{figure}

Fortunately, the simple compound Poisson model allows not only for a solution, but the reweighting can also be treated analytically.
From the technical point of view, there are two approaches to extracting the expectations of interest from the computer-generated realizations of the model. The first, to which we refer to as ``path-based'', calculates each quantity of interest for all generated paths, and afterwards calculates the statistics of the gained datasets. The second, to which we refer to as ``surface-based'', estimates the time-dependent probability density function of the portfolio loss, i.\ e., a simple matrix, see Fig.\ \ref{fig:losssurf} . This work deals with the former approach since it is more suitable for our purposes, which is getting a better estimate of the price of the more senior tranches, i.\ e., to reduce the number of Monte Carlo paths necessary for obtaining a certain precision.

\subsection{Monte Carlo with paths}

The $i$th realization of the process
is translated to financial values taking into account only the cash flows:
\begin{align}
\label{eq:idefpath} \Xdef_i & = \sum_{k=1}^{k_{\max}} e^{-rt_k} \left(\ell^{a,d}_{t_k}-\ell^{a,d}_{t_{k-1}}\right)\\
\label{eq:iprempath} \Xprem_i & = \sum_{k=1}^{k_{\max}} e^{-rt_k}\, \ON_{t_k}, 
\end{align}
with $0 = t_0, t_1, \ldots, t_{k_{\max}} = M$ representing an equally spaced time grid.
The necessity of discretization is commonly a drawback of simulations, in our case, however, it is nearer to a real world scenario since CDO payments are typically transferred quarterly.
For efficiency reasons, the generated paths should be evaluated simultaneously for all
tranches of interest.
The individual financial values $\Xdef_i$ and $\Xprem_i$ per tranche are subject to a statistical analysis.

\subsection{Reweighted Monte Carlo}

The aim of reweighting is to reduce the computational time needed for finding the expected present values of the premium and the default.
Applying reweighting, the program generates Monte Carlo paths with an altered parameter set of the compound Poisson model, that is, with an altered intensity $\rho\Alt$ and an altered expected jump size $\frac{1}{\lambda\Alt}$.
Using parameters which describe a relatively calm economic situation, senior tranches are typically not reached by MC paths, i.\ e., one obtains a poor estimate of their price. The idea is to simulate paths using a parameter set describing a more severe situation, and reweight them to preserve the original expected value while reducing the variance.

The reweighting relies on the \RN derivative, the derivative of one probability measure with respect to another \cite{Staum2003a}.
To get the appropriate weights, we have to calculate it for the real with respect to the altered one.

We write the probability of realization of a path considering that the $N_M$ jumps occur in small intervals $(t_j,t_j+\ud t)$ $j=1,\dots,N_M$ with the given intensity $\rho$ and that there is no jump outside of these intervals.
We handle the jump sizes in a similar way under the condition that the number of jumps is already given by the Poisson process (we introduce the notations $t_0=0$ and $D_0=0$):
\begin{multline}\label{eq:ppath}
\P(\mathrm{path}) =
    \prod_{i=1}^{N_M} \left(e^{-\rho(t_i-t_{i-1})}\rho \,\ud t\right) \cdot
    e^{-\rho(M-t_{N_M})} \\
    \times \prod_{i=1}^{N_M} \left(e^{-\lambda(D_{t_i}-D_{t_{i-1}})}\lambda \,\ud h\right),
\end{multline}
with the first term corresponding to the pdf of the intervals between jumps, the second to the probability that there is no event between the last jump and the maturity $M$ and the third to the pdf of the jump sizes.
It is straightforward to generalize the latter equation for arbitrary renewal processes, one only has to change the terms to the jump time and size distributions of interest.

The \RN derivative of one probability measure with respect to another one is the ratio of the weights of the same path given in \eqref{eq:ppath} under the two measures, and for a path with $N_M$ jumps is given by
\begin{equation}\begin{aligned}
&R(N_M,D_M)\\
&\quad=\frac{\ud\P\Rel}{\ud\P\Alt}(N_M,D_M) \\
&\quad=\left(\frac{\rho\Rel\lambda\Rel}{\rho\Alt\lambda\Alt}\right)^{N_M}
e^{-(\rho\Rel-\rho\Alt)M-(\lambda\Rel-\lambda\Alt)D_M}
\end{aligned}\label{eq:rnder}\end{equation}
where $\P\Rel$ and $\P\Alt$ are respectively the measures parametrized by the
real and the alternative parameters. This quantity is both calculable
analytically as a random variable and numerically for a specific Monte Carlo
path. The Monte Carlo simulation calculates $R(\mathrm{path})$ step by step,
during the generation of a path. Starting from the value 1, at each jump in the
path, the program multiplies the stored value by the contribution of that jump
(terms under the product signs in \eqref{eq:ppath}); at the end of the path, it
multiplies the value by the contribution which describes that no more events
happened until the maturity (middle factor on the right-hand side of
\eqref{eq:ppath}).

In mathematical terms, the random variable simulated under the altered measure $\P\Alt$ is $RX$ (with $X$ standing for either premium or default leg), thus, its observable variance is
\begin{equation}\begin{aligned}
\var\Alt(RX)&=\E\Alt(R^2X^2)-(\E\Alt RX)^2\\
&=\E\Rel(RX^2)-(\E\Rel X)^2,
\end{aligned}\label{eq:rwvar2}\end{equation}
since $\E\Alt\br{RX} = \E\br X$, by definition of the \RN derivative.

Being able to measure the variance, we use this feature to find the optimal
parameter set for the speed of convergence and then we perform importance sampling
with those parameters.

\subsection{Reweighting: Analytic approach}

In this section, we analytically evaluate the variance of the reweighted default considering the compound Poisson process $D_t$ where the jump times follow a Poisson process with intensity $\rho$ and the sizes of the jumps are independent exponentially distributed random variables with parameter $\lambda$, which are independent of the Poisson point process as well.
Then we calculate the expected loss and its variance analytically with the assumption that there are no discount factors, i.\ e., $r=0$.

Recall \eqref{eq:defXdef}, and note that in case $r=0$, we have
\begin{equation}\begin{aligned}
\Xdef =& \br{L_t-a}\cdot\I\br{L_t\in\Br{a,d}} \\
&+\br{d-a}\cdot\1\br{L_t>d}\\
=&\br{1-e^{-D_t}-a}
\cdot\I\br{D_t\in\Br{\ha,\hd}}\\
&+\br{d-a}\cdot\I\br{D_t>\hd}.
\end{aligned}\label{eq:defproc}\end{equation}
Then
\begin{equation}\label{eq:rweval}
\defPV=\E\Rel\br{\Xdef}=\E\Alt\br{R\Xdef},
\end{equation}
where the second equality holds by the definition of the measure change.
The important quantity here is the error of the Monte Carlo simulation carried out with importance sampling, thus, our aim is to calculate the variance given in \eqref{eq:rwvar2}.

The joint density of $N_M$ (number of jumps) and $D_M$ is given by
\begin{align*}
&f_*(n,h)\\
&\quad=\P_*(N_M=n,D_M\in(h,h+\ud h))\\
&\quad=e^{-\rho_*M}\,\frac{(\rho_*M)^n }{n!}\cdot h^{n-1} e^{-\lambda_*h}
\frac{(\lambda_*)^n}{(n-1)!}\,\ud h+o(\ud h)\\
&\quad = e^{-\lambda_*h-\rho_*M}\,\frac{(\rho_*M\lambda_*)^n
h^{n-1}}{n!(n-1)!}\,\ud h+o(\ud h)
\end{align*}
where $*$ can stand for either altered or real. This bivariate joint probability density function is composed of the product of the probability density functions of a $\mathrm{POI}(\rho_*M)$ describing $n$ events until the maturity and a $\Gamma(\lambda_*,n)$ describing the conditional probability of arriving in $(h, h+\ud h)$ having $n$ jumps.
Please note that this is a defective probability distribution, the missing mass is $\P_*(D_M=0)=e^{-\rho_*M}$.

Using \eqref{eq:defproc}, one can calculate
\begin{equation}\begin{aligned}
\E\Rel\left(\Xdef\right)
&=(1-a)\int_{\ha}^{\hd} \sum_{n=1}^\infty f\Rel(n,h)\,\ud h \\
&\quad-\int_{\ha}^{\hd} e^{-h}\sum_{n=1}^\infty f\Rel(n,h)\,\ud h\\
&\quad+(d-a)\int_{\hd}^\infty \sum_{n=1}^\infty f\Rel(n,h)\,\ud h,
\end{aligned}\end{equation}
similarly, for the variance,
\begin{equation}\begin{aligned}
&\E\Rel\left(R\left(\Xdef\right)^2\right)\\
&\quad=\int_{\ha}^{\hd} \sum_{n=1}^\infty e^{-2h} R(n,h) f\Rel(n,h)\,\ud h\\
&\qquad-2(1-a)\int_{\ha}^{\hd} \sum_{n=1}^\infty e^{-h} R(n,h) f\Rel(n,h)\,\ud h\\
&\qquad+(1-a)^2\int_{\ha}^{\hd} \sum_{n=1}^\infty R(n,h) f\Rel(n,h)\,\ud h\\
&\qquad+(d-a)^2 \int_{\hd}^\infty \sum_{n=1}^\infty R(n,h) f\Rel(n,h)\,\ud h\\
\end{aligned}\label{eq:ERXX}\end{equation}
where the integrals can be expressed in terms of incomplete gamma functions,
since the dependence of the integrands on $h$ is a product of a polynomial and
an exponential function, that is, they are of the form
\[\int_l^u e^{-\nu h}h^{n-1}\,\d h.\]
Deriving the result is not extremely difficult but rather technical, therefore we omit these details. 

We remark one more interesting feature of the expectation in \eqref{eq:ERXX}
which is the appearance of \emph{phase transition}. In the first two terms on
the right-hand side of \eqref{eq:ERXX}, the integrals are not necessarily
finite (for the other term, we do not have this issue). It can be verified by
analyzing the exponential factors in $h$ of the integrand. In the first term on
the right-hand side of \eqref{eq:ERXX}, $R(n,h)$ contributes with
$e^{-(\lambda\Rel-\lambda\Alt)h}$, whereas $f(n,h)$ gives an exponential factor
of $e^{-\lambda\Rel h}$. The product of these two is clearly
$e^{-(2\lambda\Rel-\lambda\Alt)h}$. Hence, the first integral in
\eqref{eq:ERXX} is finite if and only if
\[2\lambda\Rel-\lambda\Alt>0\quad\iff\quad\mu\Alt>\frac12\,\mu\Rel,\]
but it does not mean a restriction in the practical point of view, because as mentioned earlier, one can only expect an improvement in the variance in case it is more likely to reach a senior tranche under $\P\Alt$ than under $\P$. (For the finiteness of the second integral, a weaker condition is sufficient.)


\subsection{Differences between the Monte Carlo and the analytic method}

We emphasize here that the difference between the analytic and the Monte Carlo
methods is not merely technical (e.g.\ discretization), but conceptual as well.
The analytic method gives all quantities of interest for a given input
parameter set (if it corresponds to an analytically solvable model), namely, we
get immediately expected value and variance for a given single tranche.

In contrast, the Monte Carlo simulation has to generate a given number of simulations divided into packages (or not) to give the financial expected values $\defPV$ and $\premPVbp$ for a chosen tranche, furthermore, doing statistics on these experimental financial values provides information about their variance.
At this point the Monte Carlo method shows an advantage.
Having all these realizations, we can simultaneously calculate the expected values and variances for any chosen set of tranches, without generating new paths.
Hence, for efficient measurement, one has to define a set of observed tranches and query the financial values simultaneously.

As pointed out several times, besides the design differences, the Monte Carlo method provides a powerful tool in cases where no analytic solution can be found.
Computers can still model and capture complex processes generating realizations, even if there are several specialties in the contract.

\subsection{Monte Carlo results}

In this section we discuss the simulation results and compare them to the analytic solution.
For illustration purposes, we have chosen from the standard tranches (see Section \ref{sec:Introduction}) the \emph{super senior tranche} ($a=0.3$, $d=1$).
For each Monte Carlo simulation we used $10^6$ paths and no recovery ($\Rec=0$).

Approximating the model parameters in a calm economic situation by:
\begin{equation}\label{par:1}
\rho\Rel=0.05~\frac{\text{events}}{\text{year}}
\end{equation}
for the intensity of the compound Poisson process and about
\begin{equation}\label{par:2}
\frac{1}{\lambda\Rel}=0.10~\frac{\text{part of the original notional}}{\text{event}}
\end{equation}
for the expected number of defaults (i.e., 10 assets are expected to default per event for a 100-element portfolio). The standard maturity is
\begin{equation}
\label{par:3} M = 5~\text{years}.
\end{equation}

The gain of the reweighting procedure is given by the well-known fact that the standard deviation of a sample average (of independent realizations of a random variable) is $\frac{\sigma}{\sqrt{n}}$, where $\sigma$ is the standard deviation of the random variable and $n$ is the number of realizations used to estimate its mean-value.
So, for the same precision, whatever we gain on $\sigma$ the square of it is gained on the number $n$ of Monte Carlo paths to be generated
\begin{equation}\label{eq:gnum}
G_\mathrm{num} := \frac{n\Rel}{n\Alt} = \frac{\sigma\Rel^2}{\sigma\Alt^2}.
\end{equation}

The real-world parameters predict relatively rare events (one event forecast for four realizations) and small chance to touch the super senior tranche.
Thus, one can expect the largest decrease in $\sigma$ for the super senior tranche when the alternative parameters $\rho\Alt$ and $\frac{1}{\lambda\Alt}$ are chosen to be larger than the real ones which associates more trials to the super senior tranche without neglecting the others.

\begin{figure}[htbp]
\begingroup
  \makeatletter
  \providecommand\color[2][]{%
    \GenericError{(gnuplot) \space\space\space\@spaces}{%
      Package color not loaded in conjunction with
      terminal option `colourtext'%
    }{See the gnuplot documentation for explanation.%
    }{Either use 'blacktext' in gnuplot or load the package
      color.sty in LaTeX.}%
    \renewcommand\color[2][]{}%
  }%
  \providecommand\includegraphics[2][]{%
    \GenericError{(gnuplot) \space\space\space\@spaces}{%
      Package graphicx or graphics not loaded%
    }{See the gnuplot documentation for explanation.%
    }{The gnuplot epslatex terminal needs graphicx.sty or graphics.sty.}%
    \renewcommand\includegraphics[2][]{}%
  }%
  \providecommand\rotatebox[2]{#2}%
  \@ifundefined{ifGPcolor}{%
    \newif\ifGPcolor
    \GPcolortrue
  }{}%
  \@ifundefined{ifGPblacktext}{%
    \newif\ifGPblacktext
    \GPblacktexttrue
  }{}%
  \let\gplgaddtomacro\g@addto@macro
  \gdef\gplbacktext{}%
  \gdef\gplfronttext{}%
  \makeatother
  \ifGPblacktext
    \def\colorrgb#1{}%
    \def\colorgray#1{}%
  \else
    \ifGPcolor
      \def\colorrgb#1{\color[rgb]{#1}}%
      \def\colorgray#1{\color[gray]{#1}}%
      \expandafter\def\csname LTw\endcsname{\color{white}}%
      \expandafter\def\csname LTb\endcsname{\color{black}}%
      \expandafter\def\csname LTa\endcsname{\color{black}}%
      \expandafter\def\csname LT0\endcsname{\color[rgb]{1,0,0}}%
      \expandafter\def\csname LT1\endcsname{\color[rgb]{0,1,0}}%
      \expandafter\def\csname LT2\endcsname{\color[rgb]{0,0,1}}%
      \expandafter\def\csname LT3\endcsname{\color[rgb]{1,0,1}}%
      \expandafter\def\csname LT4\endcsname{\color[rgb]{0,1,1}}%
      \expandafter\def\csname LT5\endcsname{\color[rgb]{1,1,0}}%
      \expandafter\def\csname LT6\endcsname{\color[rgb]{0,0,0}}%
      \expandafter\def\csname LT7\endcsname{\color[rgb]{1,0.3,0}}%
      \expandafter\def\csname LT8\endcsname{\color[rgb]{0.5,0.5,0.5}}%
    \else
      \def\colorrgb#1{\color{black}}%
      \def\colorgray#1{\color[gray]{#1}}%
      \expandafter\def\csname LTw\endcsname{\color{white}}%
      \expandafter\def\csname LTb\endcsname{\color{black}}%
      \expandafter\def\csname LTa\endcsname{\color{black}}%
      \expandafter\def\csname LT0\endcsname{\color{black}}%
      \expandafter\def\csname LT1\endcsname{\color{black}}%
      \expandafter\def\csname LT2\endcsname{\color{black}}%
      \expandafter\def\csname LT3\endcsname{\color{black}}%
      \expandafter\def\csname LT4\endcsname{\color{black}}%
      \expandafter\def\csname LT5\endcsname{\color{black}}%
      \expandafter\def\csname LT6\endcsname{\color{black}}%
      \expandafter\def\csname LT7\endcsname{\color{black}}%
      \expandafter\def\csname LT8\endcsname{\color{black}}%
    \fi
  \fi
  \setlength{\unitlength}{0.0500bp}%
  \begin{picture}(4320.00,3022.00)%
    \gplgaddtomacro\gplbacktext{%
      \csname LTb\endcsname%
      \put(860,640){\makebox(0,0)[r]{\strut{}   0}}%
      \put(860,925){\makebox(0,0)[r]{\strut{} 100}}%
      \put(860,1210){\makebox(0,0)[r]{\strut{} 200}}%
      \put(980,440){\makebox(0,0){\strut{}0.0}}%
      \put(1576,440){\makebox(0,0){\strut{}0.2}}%
      \put(2172,440){\makebox(0,0){\strut{}0.4}}%
      \put(2767,440){\makebox(0,0){\strut{}0.6}}%
      \put(3363,440){\makebox(0,0){\strut{}0.8}}%
      \put(3959,440){\makebox(0,0){\strut{}1.0}}%
      \put(160,925){\rotatebox{-270}{\makebox(0,0){\strut{}$\D(\Xdef)$}}}%
      \put(2469,140){\makebox(0,0){\strut{}$\frac{1}{\lambda\Alt}~\left[\frac{1}{\mathrm{event}}\right]$}}%
    }%
    \gplgaddtomacro\gplfronttext{%
    }%
    \gplgaddtomacro\gplbacktext{%
      \csname LTb\endcsname%
      \put(860,1348){\makebox(0,0)[r]{\strut{} 6.2}}%
      \put(860,1826){\makebox(0,0)[r]{\strut{} 6.4}}%
      \put(860,2303){\makebox(0,0)[r]{\strut{} 6.6}}%
      \put(860,2781){\makebox(0,0)[r]{\strut{} 6.8}}%
      \put(980,1148){\makebox(0,0){\strut{}}}%
      \put(1576,1148){\makebox(0,0){\strut{}}}%
      \put(2172,1148){\makebox(0,0){\strut{}}}%
      \put(2767,1148){\makebox(0,0){\strut{}}}%
      \put(3363,1148){\makebox(0,0){\strut{}}}%
      \put(3959,1148){\makebox(0,0){\strut{}}}%
      \put(160,2064){\rotatebox{-270}{\makebox(0,0){\strut{}$\E(\Xdef)$~[BP]}}}%
      \put(2469,1088){\makebox(0,0){\strut{}}}%
    }%
    \gplgaddtomacro\gplfronttext{%
      \csname LTb\endcsname%
      \put(3056,1711){\makebox(0,0)[r]{\strut{}MC avg. of $\Xdef$}}%
      \csname LTb\endcsname%
      \put(3056,1511){\makebox(0,0)[r]{\strut{}$\E({\Xdef})\pm\mathrm{std.dev.}$}}%
    }%
    \gplbacktext
    \put(0,0){\includegraphics{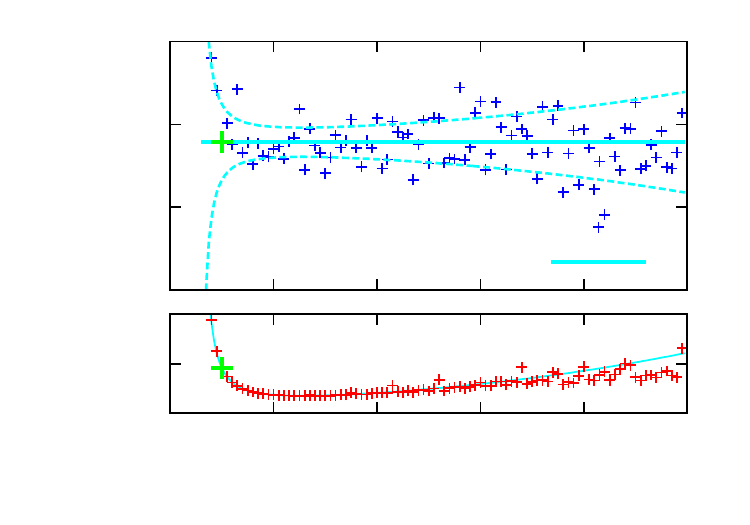}}%
    \gplfronttext
  \end{picture}%
\endgroup
\caption[$\Xdef$ as a function of $\frac{1}{\lambda\Alt}$]
{The measured average and standard deviation of $\Xdef$ for the super senior tranche as a function of $\frac{1}{\lambda\Alt}$, with $\rho\Alt = \rho\Rel$, ($\lambda\Rel = 10$, $\rho\Rel = 0.05$). The average and standard deviation of $\Xdef$ is given in basis points. The variance remarkably decreases for $\frac{1}{\lambda\Alt}\approx 2.8\times\frac{1}{\lambda\Rel}=0.28$. \coloronline}
\label{fig:dlambda15}
\end{figure}

\begin{figure}[htbp]
\begin{small}{
\begingroup
  \makeatletter
  \providecommand\color[2][]{%
    \GenericError{(gnuplot) \space\space\space\@spaces}{%
      Package color not loaded in conjunction with
      terminal option `colourtext'%
    }{See the gnuplot documentation for explanation.%
    }{Either use 'blacktext' in gnuplot or load the package
      color.sty in LaTeX.}%
    \renewcommand\color[2][]{}%
  }%
  \providecommand\includegraphics[2][]{%
    \GenericError{(gnuplot) \space\space\space\@spaces}{%
      Package graphicx or graphics not loaded%
    }{See the gnuplot documentation for explanation.%
    }{The gnuplot epslatex terminal needs graphicx.sty or graphics.sty.}%
    \renewcommand\includegraphics[2][]{}%
  }%
  \providecommand\rotatebox[2]{#2}%
  \@ifundefined{ifGPcolor}{%
    \newif\ifGPcolor
    \GPcolortrue
  }{}%
  \@ifundefined{ifGPblacktext}{%
    \newif\ifGPblacktext
    \GPblacktexttrue
  }{}%
  \let\gplgaddtomacro\g@addto@macro
  \gdef\gplbacktext{}%
  \gdef\gplfronttext{}%
  \makeatother
  \ifGPblacktext
    \def\colorrgb#1{}%
    \def\colorgray#1{}%
  \else
    \ifGPcolor
      \def\colorrgb#1{\color[rgb]{#1}}%
      \def\colorgray#1{\color[gray]{#1}}%
      \expandafter\def\csname LTw\endcsname{\color{white}}%
      \expandafter\def\csname LTb\endcsname{\color{black}}%
      \expandafter\def\csname LTa\endcsname{\color{black}}%
      \expandafter\def\csname LT0\endcsname{\color[rgb]{1,0,0}}%
      \expandafter\def\csname LT1\endcsname{\color[rgb]{0,1,0}}%
      \expandafter\def\csname LT2\endcsname{\color[rgb]{0,0,1}}%
      \expandafter\def\csname LT3\endcsname{\color[rgb]{1,0,1}}%
      \expandafter\def\csname LT4\endcsname{\color[rgb]{0,1,1}}%
      \expandafter\def\csname LT5\endcsname{\color[rgb]{1,1,0}}%
      \expandafter\def\csname LT6\endcsname{\color[rgb]{0,0,0}}%
      \expandafter\def\csname LT7\endcsname{\color[rgb]{1,0.3,0}}%
      \expandafter\def\csname LT8\endcsname{\color[rgb]{0.5,0.5,0.5}}%
    \else
      \def\colorrgb#1{\color{black}}%
      \def\colorgray#1{\color[gray]{#1}}%
      \expandafter\def\csname LTw\endcsname{\color{white}}%
      \expandafter\def\csname LTb\endcsname{\color{black}}%
      \expandafter\def\csname LTa\endcsname{\color{black}}%
      \expandafter\def\csname LT0\endcsname{\color{black}}%
      \expandafter\def\csname LT1\endcsname{\color{black}}%
      \expandafter\def\csname LT2\endcsname{\color{black}}%
      \expandafter\def\csname LT3\endcsname{\color{black}}%
      \expandafter\def\csname LT4\endcsname{\color{black}}%
      \expandafter\def\csname LT5\endcsname{\color{black}}%
      \expandafter\def\csname LT6\endcsname{\color{black}}%
      \expandafter\def\csname LT7\endcsname{\color{black}}%
      \expandafter\def\csname LT8\endcsname{\color{black}}%
    \fi
  \fi
  \setlength{\unitlength}{0.0500bp}%
  \begin{picture}(4320.00,3022.00)%
    \gplgaddtomacro\gplbacktext{%
      \csname LTb\endcsname%
      \put(860,640){\makebox(0,0)[r]{\strut{}   0}}%
      \put(860,830){\makebox(0,0)[r]{\strut{}   2}}%
      \put(860,1020){\makebox(0,0)[r]{\strut{}   4}}%
      \put(860,1210){\makebox(0,0)[r]{\strut{}   6}}%
      \put(980,440){\makebox(0,0){\strut{}0.0}}%
      \put(1576,440){\makebox(0,0){\strut{}0.2}}%
      \put(2172,440){\makebox(0,0){\strut{}0.4}}%
      \put(2767,440){\makebox(0,0){\strut{}0.6}}%
      \put(3363,440){\makebox(0,0){\strut{}0.8}}%
      \put(3959,440){\makebox(0,0){\strut{}1.0}}%
      \put(160,925){\rotatebox{-270}{\makebox(0,0){\strut{}$\D(\Xprem)$}}}%
      \put(2469,140){\makebox(0,0){\strut{}$\frac{1}{\lambda\Alt}~\left[\frac{1}{\mathrm{event}}\right]$}}%
    }%
    \gplgaddtomacro\gplfronttext{%
    }%
    \gplgaddtomacro\gplbacktext{%
      \csname LTb\endcsname%
      \put(860,1348){\makebox(0,0)[r]{\strut{}3.48}}%
      \put(860,1826){\makebox(0,0)[r]{\strut{}3.49}}%
      \put(860,2303){\makebox(0,0)[r]{\strut{}3.50}}%
      \put(860,2781){\makebox(0,0)[r]{\strut{}3.51}}%
      \put(980,1148){\makebox(0,0){\strut{}}}%
      \put(1576,1148){\makebox(0,0){\strut{}}}%
      \put(2172,1148){\makebox(0,0){\strut{}}}%
      \put(2767,1148){\makebox(0,0){\strut{}}}%
      \put(3363,1148){\makebox(0,0){\strut{}}}%
      \put(3959,1148){\makebox(0,0){\strut{}}}%
      \put(160,2064){\rotatebox{-270}{\makebox(0,0){\strut{}$\E(\Xprem)$~[BP]}}}%
      \put(2469,1088){\makebox(0,0){\strut{}}}%
    }%
    \gplgaddtomacro\gplfronttext{%
      \csname LTb\endcsname%
      \put(3056,1711){\makebox(0,0)[r]{\strut{}MC avg. for $\Xprem$}}%
      \csname LTb\endcsname%
      \put(3056,1511){\makebox(0,0)[r]{\strut{}$\E(\Xprem))$}}%
    }%
    \gplbacktext
    \put(0,0){\includegraphics{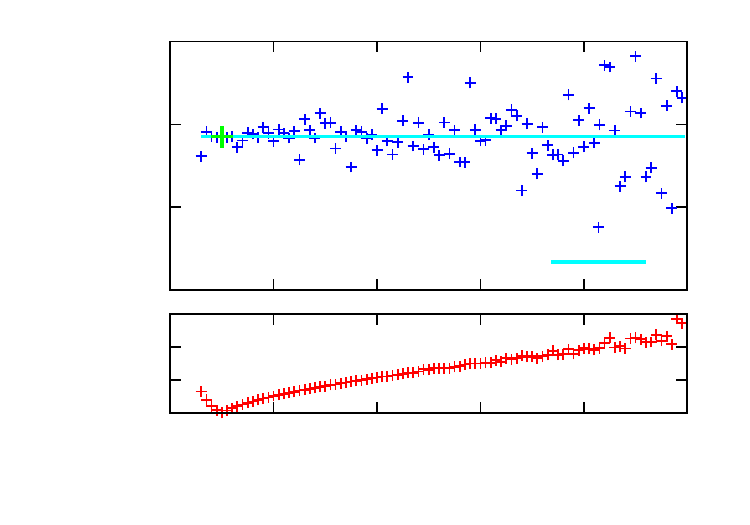}}%
    \gplfronttext
  \end{picture}%
\endgroup
}\end{small}
\caption[$\Xprem$ with regard to the reweighting parameter $\frac{1}{\lambda\Alt}$]
{The measured average and standard deviation of $\Xprem$ for the super senior tranche as a function of $\frac{1}{\lambda\Alt}$, with $\rho\Alt = \rho\Rel$, ($\lambda\Rel = 10$, $\rho\Rel = 0.05$). The average and standard deviation of $\Xprem$ is given for a 1 basis point spread.
The estimation of the premium leg present value was not improved for the super senior tranche. We have no analytic formula for $\premPVbp$. \coloronline}
\label{fig:plambda15}
\end{figure}

\begin{figure}[htbp]
\begin{small}{
\begingroup
  \makeatletter
  \providecommand\color[2][]{%
    \GenericError{(gnuplot) \space\space\space\@spaces}{%
      Package color not loaded in conjunction with
      terminal option `colourtext'%
    }{See the gnuplot documentation for explanation.%
    }{Either use 'blacktext' in gnuplot or load the package
      color.sty in LaTeX.}%
    \renewcommand\color[2][]{}%
  }%
  \providecommand\includegraphics[2][]{%
    \GenericError{(gnuplot) \space\space\space\@spaces}{%
      Package graphicx or graphics not loaded%
    }{See the gnuplot documentation for explanation.%
    }{The gnuplot epslatex terminal needs graphicx.sty or graphics.sty.}%
    \renewcommand\includegraphics[2][]{}%
  }%
  \providecommand\rotatebox[2]{#2}%
  \@ifundefined{ifGPcolor}{%
    \newif\ifGPcolor
    \GPcolortrue
  }{}%
  \@ifundefined{ifGPblacktext}{%
    \newif\ifGPblacktext
    \GPblacktexttrue
  }{}%
  \let\gplgaddtomacro\g@addto@macro
  \gdef\gplbacktext{}%
  \gdef\gplfronttext{}%
  \makeatother
  \ifGPblacktext
    \def\colorrgb#1{}%
    \def\colorgray#1{}%
  \else
    \ifGPcolor
      \def\colorrgb#1{\color[rgb]{#1}}%
      \def\colorgray#1{\color[gray]{#1}}%
      \expandafter\def\csname LTw\endcsname{\color{white}}%
      \expandafter\def\csname LTb\endcsname{\color{black}}%
      \expandafter\def\csname LTa\endcsname{\color{black}}%
      \expandafter\def\csname LT0\endcsname{\color[rgb]{1,0,0}}%
      \expandafter\def\csname LT1\endcsname{\color[rgb]{0,1,0}}%
      \expandafter\def\csname LT2\endcsname{\color[rgb]{0,0,1}}%
      \expandafter\def\csname LT3\endcsname{\color[rgb]{1,0,1}}%
      \expandafter\def\csname LT4\endcsname{\color[rgb]{0,1,1}}%
      \expandafter\def\csname LT5\endcsname{\color[rgb]{1,1,0}}%
      \expandafter\def\csname LT6\endcsname{\color[rgb]{0,0,0}}%
      \expandafter\def\csname LT7\endcsname{\color[rgb]{1,0.3,0}}%
      \expandafter\def\csname LT8\endcsname{\color[rgb]{0.5,0.5,0.5}}%
    \else
      \def\colorrgb#1{\color{black}}%
      \def\colorgray#1{\color[gray]{#1}}%
      \expandafter\def\csname LTw\endcsname{\color{white}}%
      \expandafter\def\csname LTb\endcsname{\color{black}}%
      \expandafter\def\csname LTa\endcsname{\color{black}}%
      \expandafter\def\csname LT0\endcsname{\color{black}}%
      \expandafter\def\csname LT1\endcsname{\color{black}}%
      \expandafter\def\csname LT2\endcsname{\color{black}}%
      \expandafter\def\csname LT3\endcsname{\color{black}}%
      \expandafter\def\csname LT4\endcsname{\color{black}}%
      \expandafter\def\csname LT5\endcsname{\color{black}}%
      \expandafter\def\csname LT6\endcsname{\color{black}}%
      \expandafter\def\csname LT7\endcsname{\color{black}}%
      \expandafter\def\csname LT8\endcsname{\color{black}}%
    \fi
  \fi
  \setlength{\unitlength}{0.0500bp}%
  \begin{picture}(4320.00,3022.00)%
    \gplgaddtomacro\gplbacktext{%
      \csname LTb\endcsname%
      \put(860,640){\makebox(0,0)[r]{\strut{}  40}}%
      \put(860,830){\makebox(0,0)[r]{\strut{}  80}}%
      \put(860,1020){\makebox(0,0)[r]{\strut{} 120}}%
      \put(860,1210){\makebox(0,0)[r]{\strut{} 160}}%
      \put(980,440){\makebox(0,0){\strut{}0.0}}%
      \put(1576,440){\makebox(0,0){\strut{}0.1}}%
      \put(2172,440){\makebox(0,0){\strut{}0.2}}%
      \put(2767,440){\makebox(0,0){\strut{}0.3}}%
      \put(3363,440){\makebox(0,0){\strut{}0.4}}%
      \put(3959,440){\makebox(0,0){\strut{}0.5}}%
      \put(160,925){\rotatebox{-270}{\makebox(0,0){\strut{}$\D(\Xdef)$}}}%
      \put(2469,140){\makebox(0,0){\strut{}$\rho\Alt~\left[\frac{\mathrm{events}}{\mathrm{year}}\right]$}}%
    }%
    \gplgaddtomacro\gplfronttext{%
    }%
    \gplgaddtomacro\gplbacktext{%
      \csname LTb\endcsname%
      \put(860,1348){\makebox(0,0)[r]{\strut{} 6.2}}%
      \put(860,1587){\makebox(0,0)[r]{\strut{} 6.3}}%
      \put(860,1826){\makebox(0,0)[r]{\strut{} 6.4}}%
      \put(860,2064){\makebox(0,0)[r]{\strut{} 6.5}}%
      \put(860,2303){\makebox(0,0)[r]{\strut{} 6.6}}%
      \put(860,2542){\makebox(0,0)[r]{\strut{} 6.7}}%
      \put(860,2781){\makebox(0,0)[r]{\strut{} 6.8}}%
      \put(980,1148){\makebox(0,0){\strut{}}}%
      \put(1576,1148){\makebox(0,0){\strut{}}}%
      \put(2172,1148){\makebox(0,0){\strut{}}}%
      \put(2767,1148){\makebox(0,0){\strut{}}}%
      \put(3363,1148){\makebox(0,0){\strut{}}}%
      \put(3959,1148){\makebox(0,0){\strut{}}}%
      \put(160,2064){\rotatebox{-270}{\makebox(0,0){\strut{}$\E(\Xdef)$~[BP]}}}%
      \put(2469,1088){\makebox(0,0){\strut{}}}%
    }%
    \gplgaddtomacro\gplfronttext{%
      \csname LTb\endcsname%
      \put(3056,1711){\makebox(0,0)[r]{\strut{}MC avg. of $\Xdef$}}%
      \csname LTb\endcsname%
      \put(3056,1511){\makebox(0,0)[r]{\strut{}$\E({\Xdef})\pm\mathrm{std.dev.}$}}%
    }%
    \gplbacktext
    \put(0,0){\includegraphics{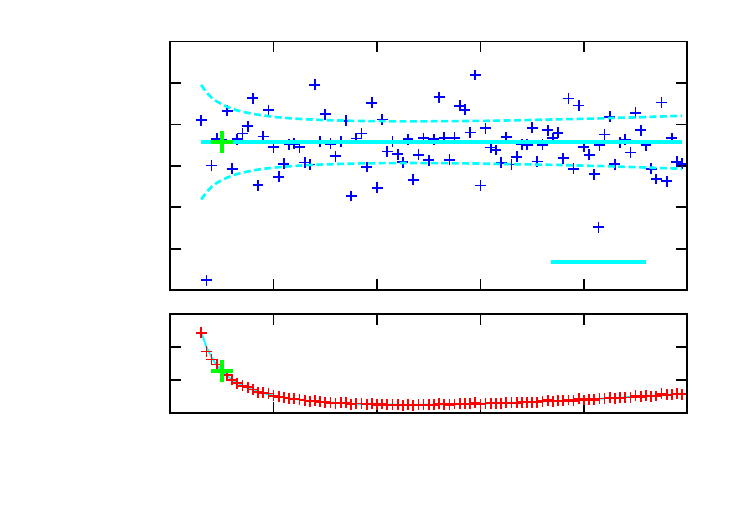}}%
    \gplfronttext
  \end{picture}%
\endgroup
}\end{small}
\caption[$\Xdef$ with regard to the reweighting parameter $\rho\Alt$]{The measured average and standard deviation of $\Xdef$ for the super senior tranche as a function of $\rho\Alt$, with $\lambda\Alt = \lambda\Rel$, ($\lambda\Rel = 10$, $\rho\Rel = 0.05$). The average and standard deviation of the $\Xdef$ is given in basis points. The variance decreases most if $\rho\Alt\in (4\rho\Rel,5\rho\Rel)=(0.20,0.25)$. \coloronline}
\label{fig:drho15}
\end{figure}

\begin{figure}[htbp]
\begin{small}{
\begingroup
  \makeatletter
  \providecommand\color[2][]{%
    \GenericError{(gnuplot) \space\space\space\@spaces}{%
      Package color not loaded in conjunction with
      terminal option `colourtext'%
    }{See the gnuplot documentation for explanation.%
    }{Either use 'blacktext' in gnuplot or load the package
      color.sty in LaTeX.}%
    \renewcommand\color[2][]{}%
  }%
  \providecommand\includegraphics[2][]{%
    \GenericError{(gnuplot) \space\space\space\@spaces}{%
      Package graphicx or graphics not loaded%
    }{See the gnuplot documentation for explanation.%
    }{The gnuplot epslatex terminal needs graphicx.sty or graphics.sty.}%
    \renewcommand\includegraphics[2][]{}%
  }%
  \providecommand\rotatebox[2]{#2}%
  \@ifundefined{ifGPcolor}{%
    \newif\ifGPcolor
    \GPcolortrue
  }{}%
  \@ifundefined{ifGPblacktext}{%
    \newif\ifGPblacktext
    \GPblacktexttrue
  }{}%
  \let\gplgaddtomacro\g@addto@macro
  \gdef\gplbacktext{}%
  \gdef\gplfronttext{}%
  \makeatother
  \ifGPblacktext
    \def\colorrgb#1{}%
    \def\colorgray#1{}%
  \else
    \ifGPcolor
      \def\colorrgb#1{\color[rgb]{#1}}%
      \def\colorgray#1{\color[gray]{#1}}%
      \expandafter\def\csname LTw\endcsname{\color{white}}%
      \expandafter\def\csname LTb\endcsname{\color{black}}%
      \expandafter\def\csname LTa\endcsname{\color{black}}%
      \expandafter\def\csname LT0\endcsname{\color[rgb]{1,0,0}}%
      \expandafter\def\csname LT1\endcsname{\color[rgb]{0,1,0}}%
      \expandafter\def\csname LT2\endcsname{\color[rgb]{0,0,1}}%
      \expandafter\def\csname LT3\endcsname{\color[rgb]{1,0,1}}%
      \expandafter\def\csname LT4\endcsname{\color[rgb]{0,1,1}}%
      \expandafter\def\csname LT5\endcsname{\color[rgb]{1,1,0}}%
      \expandafter\def\csname LT6\endcsname{\color[rgb]{0,0,0}}%
      \expandafter\def\csname LT7\endcsname{\color[rgb]{1,0.3,0}}%
      \expandafter\def\csname LT8\endcsname{\color[rgb]{0.5,0.5,0.5}}%
    \else
      \def\colorrgb#1{\color{black}}%
      \def\colorgray#1{\color[gray]{#1}}%
      \expandafter\def\csname LTw\endcsname{\color{white}}%
      \expandafter\def\csname LTb\endcsname{\color{black}}%
      \expandafter\def\csname LTa\endcsname{\color{black}}%
      \expandafter\def\csname LT0\endcsname{\color{black}}%
      \expandafter\def\csname LT1\endcsname{\color{black}}%
      \expandafter\def\csname LT2\endcsname{\color{black}}%
      \expandafter\def\csname LT3\endcsname{\color{black}}%
      \expandafter\def\csname LT4\endcsname{\color{black}}%
      \expandafter\def\csname LT5\endcsname{\color{black}}%
      \expandafter\def\csname LT6\endcsname{\color{black}}%
      \expandafter\def\csname LT7\endcsname{\color{black}}%
      \expandafter\def\csname LT8\endcsname{\color{black}}%
    \fi
  \fi
  \setlength{\unitlength}{0.0500bp}%
  \begin{picture}(4320.00,3022.00)%
    \gplgaddtomacro\gplbacktext{%
      \csname LTb\endcsname%
      \put(860,640){\makebox(0,0)[r]{\strut{} 0.0}}%
      \put(860,925){\makebox(0,0)[r]{\strut{} 5.0}}%
      \put(860,1210){\makebox(0,0)[r]{\strut{}10.0}}%
      \put(980,440){\makebox(0,0){\strut{}0.0}}%
      \put(1576,440){\makebox(0,0){\strut{}0.1}}%
      \put(2172,440){\makebox(0,0){\strut{}0.2}}%
      \put(2767,440){\makebox(0,0){\strut{}0.3}}%
      \put(3363,440){\makebox(0,0){\strut{}0.4}}%
      \put(3959,440){\makebox(0,0){\strut{}0.5}}%
      \put(160,925){\rotatebox{-270}{\makebox(0,0){\strut{}$\D(\Xprem)$}}}%
      \put(2469,140){\makebox(0,0){\strut{}$\rho\Alt~\left[\frac{\mathrm{events}}{\mathrm{year}}\right]$}}%
    }%
    \gplgaddtomacro\gplfronttext{%
    }%
    \gplgaddtomacro\gplbacktext{%
      \csname LTb\endcsname%
      \put(860,1348){\makebox(0,0)[r]{\strut{}3.48}}%
      \put(860,2065){\makebox(0,0)[r]{\strut{}3.50}}%
      \put(860,2781){\makebox(0,0)[r]{\strut{}3.52}}%
      \put(980,1148){\makebox(0,0){\strut{}}}%
      \put(1576,1148){\makebox(0,0){\strut{}}}%
      \put(2172,1148){\makebox(0,0){\strut{}}}%
      \put(2767,1148){\makebox(0,0){\strut{}}}%
      \put(3363,1148){\makebox(0,0){\strut{}}}%
      \put(3959,1148){\makebox(0,0){\strut{}}}%
      \put(160,2064){\rotatebox{-270}{\makebox(0,0){\strut{}$\E(\Xprem)$~[BP]}}}%
      \put(2469,1088){\makebox(0,0){\strut{}}}%
    }%
    \gplgaddtomacro\gplfronttext{%
      \csname LTb\endcsname%
      \put(3056,1711){\makebox(0,0)[r]{\strut{}MC avg. for $\Xprem$}}%
      \csname LTb\endcsname%
      \put(3056,1511){\makebox(0,0)[r]{\strut{}$\E(\Xprem))$}}%
    }%
    \gplbacktext
    \put(0,0){\includegraphics{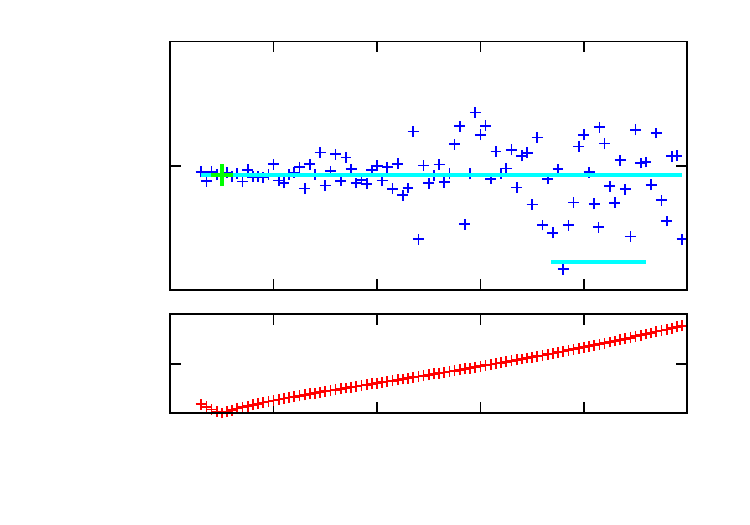}}%
    \gplfronttext
  \end{picture}%
\endgroup
}\end{small}
\caption[$\Xprem$ with regard to the reweighting parameter $\rho\Alt$]{The measured average and standard deviation of $\Xprem$ for the super senior tranche as a function of $\rho\Alt$, with $\lambda\Alt = \lambda\Rel$, ($\lambda\Rel = 10$, $\rho\Rel = 0.05$). The average and standard deviation of $\Xprem$ is given for a 1 basis point spread. The estimation of the premium leg present value was not improved for the super senior tranche. \coloronline}
\label{fig:prho15}
\end{figure}

We executed a simulation for the given $\rho\Rel$, $\frac{1}{\lambda\Rel}$ and $M=5$ to study the effect of the \emph{jump size parameter} on the standard deviation.
The results for the super senior tranche are presented in Figures \ref{fig:dlambda15} and \ref{fig:plambda15}.
The $\Xdef$ and $\Xprem$ scales are in basis points (BP), i.e., values are multiplied by $10^4$, according to financial convention.
In the top part of each diagram we represent the empirical expected value, while in the background the analytic one, together with the analytically calculated Monte Carlo error as confidence interval.
This error is given by the standard deviation of the $\Xdef$ for one path divided by $\sqrt{n}$, where $n$ is the number of the Monte Carlo paths used in the measurement.
The bottom part of the diagrams retraces the analytic and the empirical value of the same standard deviation to facilitate the comparison.
The expected gain in computational time is proportional to the decrease of the variance, that is, the square of the decrease in the standard deviation plotted here.

We observe a very good match of the analytic and numerical (path) approach, and in addition, we discover that we can realize gains in precision for the index tranche as well as for the tranches above the equity tranche.
The figures confirm our expectation that the variance of $\Xdef$ can be decreased.
The maximal gain for the super senior tranche is obtained for $\frac{1}{\lambda\Alt}=0.28$ where the variance was reduced to $14\%$ of its original value.
In contrast, $\premPVbp$ cannot be estimated more accurately than in the non-reweighted case.
It should not discourage us, because the most uncertainty of the CDO contracts originates from the wrong estimation of $\Xdef$'s properties. The properties of $\Xprem$ can already be easily and accurately estimated without reweighting as shown in the figures. Even the reweighting-increased relative error ($\frac{\text{std. dev.}}{\text{expected v.}}$) of $\premPVbp$ remains an order below the relative error of $\defPV$.
The reweighting is therefore improving the most important part of the pricing.

A similar investigation was done on the \emph{jump time parameter} (i.e., the \emph{intensity} of the compound Poisson process).
Using the same real-world parameters the results for the super senior tranche are represented in Figures \ref{fig:drho15} and \ref{fig:prho15}.

We observe an equally good match of the analytic and numerical (path) approach
as previously. Here, we are glad to observe that all the tranches exhibit gain
in precision with growing process intensity. The figures confirm our
expectations that the variance of $\Xdef$ can be decreased. The maximal gain
for the index tranche is obtained for $\rho\Alt=0.22$ where the variance was
reduced to $32\%$ of its original value.

\begin{figure}[htbp]
\begingroup
  \makeatletter
  \providecommand\color[2][]{%
    \GenericError{(gnuplot) \space\space\space\@spaces}{%
      Package color not loaded in conjunction with
      terminal option `colourtext'%
    }{See the gnuplot documentation for explanation.%
    }{Either use 'blacktext' in gnuplot or load the package
      color.sty in LaTeX.}%
    \renewcommand\color[2][]{}%
  }%
  \providecommand\includegraphics[2][]{%
    \GenericError{(gnuplot) \space\space\space\@spaces}{%
      Package graphicx or graphics not loaded%
    }{See the gnuplot documentation for explanation.%
    }{The gnuplot epslatex terminal needs graphicx.sty or graphics.sty.}%
    \renewcommand\includegraphics[2][]{}%
  }%
  \providecommand\rotatebox[2]{#2}%
  \@ifundefined{ifGPcolor}{%
    \newif\ifGPcolor
    \GPcolortrue
  }{}%
  \@ifundefined{ifGPblacktext}{%
    \newif\ifGPblacktext
    \GPblacktexttrue
  }{}%
  \let\gplgaddtomacro\g@addto@macro
  \gdef\gplbacktext{}%
  \gdef\gplfronttext{}%
  \makeatother
  \ifGPblacktext
    \def\colorrgb#1{}%
    \def\colorgray#1{}%
  \else
    \ifGPcolor
      \def\colorrgb#1{\color[rgb]{#1}}%
      \def\colorgray#1{\color[gray]{#1}}%
      \expandafter\def\csname LTw\endcsname{\color{white}}%
      \expandafter\def\csname LTb\endcsname{\color{black}}%
      \expandafter\def\csname LTa\endcsname{\color{black}}%
      \expandafter\def\csname LT0\endcsname{\color[rgb]{1,0,0}}%
      \expandafter\def\csname LT1\endcsname{\color[rgb]{0,1,0}}%
      \expandafter\def\csname LT2\endcsname{\color[rgb]{0,0,1}}%
      \expandafter\def\csname LT3\endcsname{\color[rgb]{1,0,1}}%
      \expandafter\def\csname LT4\endcsname{\color[rgb]{0,1,1}}%
      \expandafter\def\csname LT5\endcsname{\color[rgb]{1,1,0}}%
      \expandafter\def\csname LT6\endcsname{\color[rgb]{0,0,0}}%
      \expandafter\def\csname LT7\endcsname{\color[rgb]{1,0.3,0}}%
      \expandafter\def\csname LT8\endcsname{\color[rgb]{0.5,0.5,0.5}}%
    \else
      \def\colorrgb#1{\color{black}}%
      \def\colorgray#1{\color[gray]{#1}}%
      \expandafter\def\csname LTw\endcsname{\color{white}}%
      \expandafter\def\csname LTb\endcsname{\color{black}}%
      \expandafter\def\csname LTa\endcsname{\color{black}}%
      \expandafter\def\csname LT0\endcsname{\color{black}}%
      \expandafter\def\csname LT1\endcsname{\color{black}}%
      \expandafter\def\csname LT2\endcsname{\color{black}}%
      \expandafter\def\csname LT3\endcsname{\color{black}}%
      \expandafter\def\csname LT4\endcsname{\color{black}}%
      \expandafter\def\csname LT5\endcsname{\color{black}}%
      \expandafter\def\csname LT6\endcsname{\color{black}}%
      \expandafter\def\csname LT7\endcsname{\color{black}}%
      \expandafter\def\csname LT8\endcsname{\color{black}}%
    \fi
  \fi
  \setlength{\unitlength}{0.0500bp}%
  \begin{picture}(4320.00,3022.00)%
    \gplgaddtomacro\gplbacktext{%
      \csname LTb\endcsname%
      \put(814,704){\makebox(0,0)[r]{\strut{}  0}}%
      \put(814,1046){\makebox(0,0)[r]{\strut{}  5}}%
      \put(814,1388){\makebox(0,0)[r]{\strut{} 10}}%
      \put(814,1731){\makebox(0,0)[r]{\strut{} 15}}%
      \put(814,2073){\makebox(0,0)[r]{\strut{} 20}}%
      \put(814,2415){\makebox(0,0)[r]{\strut{} 25}}%
      \put(814,2757){\makebox(0,0)[r]{\strut{} 30}}%
      \put(946,484){\makebox(0,0){\strut{}0.0}}%
      \put(1442,484){\makebox(0,0){\strut{}1.0}}%
      \put(1938,484){\makebox(0,0){\strut{}2.0}}%
      \put(2435,484){\makebox(0,0){\strut{}3.0}}%
      \put(2931,484){\makebox(0,0){\strut{}4.0}}%
      \put(3427,484){\makebox(0,0){\strut{}5.0}}%
      \put(3923,484){\makebox(0,0){\strut{}6.0}}%
      \put(176,1730){\rotatebox{-270}{\makebox(0,0){\strut{}execution time [s]}}}%
      \put(2434,154){\makebox(0,0){\strut{}$\rho\Alt$ [events/year]}}%
    }%
    \gplgaddtomacro\gplfronttext{%
      \csname LTb\endcsname%
      \put(2936,1097){\makebox(0,0)[r]{\strut{}Linear fit}}%
      \csname LTb\endcsname%
      \put(2936,877){\makebox(0,0)[r]{\strut{}CPU time}}%
    }%
    \gplbacktext
    \put(0,0){\includegraphics{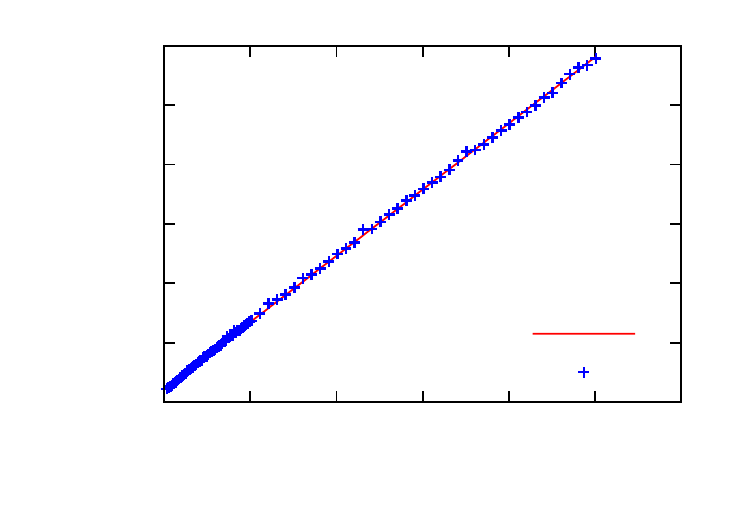}}%
    \gplfronttext
  \end{picture}%
\endgroup
\caption[The computational time as a function of the reweighting parameter $\rho\Alt$]{The computational time as a function of the reweighting parameter $\rho\Alt$, which denotes the Poisson process intensity. The points fit on a line of equation $t=c+b\rho\Alt$ where $c=1.1(67)$ is the average initialization cost of the $10^6$ MC paths and $b=5.57(1)$ is the time consumed by processing the jumps. \coloronline}
\label{fig:chrono}
\end{figure}

Although by increasing $\rho$ we can gain on the number of Monte Carlo paths, the gain on computation power is not so evident.
On average, each path contains $M \rho\Alt$ jumps which need to be generated by the random number generator and administrated by the financial layer.
An explicit chronometry was done where the process' real computational time on the CPU was collected (not the clock ticks during the execution because the computer could do other things in the background).
The time need for increased $\rho\Alt$ is plotted in Figure
\ref{fig:chrono} and it is linear in $\rho\Alt$. With regard to this fact the real gain in computational time is
\begin{equation}\label{eq:gtime}
G_\mathrm{time} = \frac{t_{\text{comp}} \Rel}{t\Alt_{\text{comp}} }
= \frac{\sigma\Rel^2}{\sigma\Alt^2} \frac{\rho \Alt}{\rho \Rel},
\end{equation}
where $\Alt$ stands for the reweighted simulation's results,
$\sigma$ denotes the standard deviation of the measured quantity ($\Xdef$ or
$\Xprem$) and $\rho$ denotes the Poisson process intensity. This effect is not
present in the case of $\frac{1}{\lambda\Alt}$ because higher jumps do not
provoke more events -- they only result in larger numeric values.
We note here that the simulation with
$n=10^6$ paths near the optimum region takes about 3 seconds for each point.

In possession of all this information, we expect that the optimum for speed will be somewhere in the region where both altered the intensity and the jump size parameter are higher than the real ones.
To find the maximal gain, we have to analyze it in two dimensions, therefore we draw a map for the gain in MC paths and the gain in time as defined respectively in \eqref{eq:gnum} and \eqref{eq:gtime}.

\begin{figure}[htbp]
{
\begingroup
  \makeatletter
  \providecommand\color[2][]{%
    \GenericError{(gnuplot) \space\space\space\@spaces}{%
      Package color not loaded in conjunction with
      terminal option `colourtext'%
    }{See the gnuplot documentation for explanation.%
    }{Either use 'blacktext' in gnuplot or load the package
      color.sty in LaTeX.}%
    \renewcommand\color[2][]{}%
  }%
  \providecommand\includegraphics[2][]{%
    \GenericError{(gnuplot) \space\space\space\@spaces}{%
      Package graphicx or graphics not loaded%
    }{See the gnuplot documentation for explanation.%
    }{The gnuplot epslatex terminal needs graphicx.sty or graphics.sty.}%
    \renewcommand\includegraphics[2][]{}%
  }%
  \providecommand\rotatebox[2]{#2}%
  \@ifundefined{ifGPcolor}{%
    \newif\ifGPcolor
    \GPcolortrue
  }{}%
  \@ifundefined{ifGPblacktext}{%
    \newif\ifGPblacktext
    \GPblacktexttrue
  }{}%
  \let\gplgaddtomacro\g@addto@macro
  \gdef\gplbacktext{}%
  \gdef\gplfronttext{}%
  \makeatother
  \ifGPblacktext
    \def\colorrgb#1{}%
    \def\colorgray#1{}%
  \else
    \ifGPcolor
      \def\colorrgb#1{\color[rgb]{#1}}%
      \def\colorgray#1{\color[gray]{#1}}%
      \expandafter\def\csname LTw\endcsname{\color{white}}%
      \expandafter\def\csname LTb\endcsname{\color{black}}%
      \expandafter\def\csname LTa\endcsname{\color{black}}%
      \expandafter\def\csname LT0\endcsname{\color[rgb]{1,0,0}}%
      \expandafter\def\csname LT1\endcsname{\color[rgb]{0,1,0}}%
      \expandafter\def\csname LT2\endcsname{\color[rgb]{0,0,1}}%
      \expandafter\def\csname LT3\endcsname{\color[rgb]{1,0,1}}%
      \expandafter\def\csname LT4\endcsname{\color[rgb]{0,1,1}}%
      \expandafter\def\csname LT5\endcsname{\color[rgb]{1,1,0}}%
      \expandafter\def\csname LT6\endcsname{\color[rgb]{0,0,0}}%
      \expandafter\def\csname LT7\endcsname{\color[rgb]{1,0.3,0}}%
      \expandafter\def\csname LT8\endcsname{\color[rgb]{0.5,0.5,0.5}}%
    \else
      \def\colorrgb#1{\color{black}}%
      \def\colorgray#1{\color[gray]{#1}}%
      \expandafter\def\csname LTw\endcsname{\color{white}}%
      \expandafter\def\csname LTb\endcsname{\color{black}}%
      \expandafter\def\csname LTa\endcsname{\color{black}}%
      \expandafter\def\csname LT0\endcsname{\color{black}}%
      \expandafter\def\csname LT1\endcsname{\color{black}}%
      \expandafter\def\csname LT2\endcsname{\color{black}}%
      \expandafter\def\csname LT3\endcsname{\color{black}}%
      \expandafter\def\csname LT4\endcsname{\color{black}}%
      \expandafter\def\csname LT5\endcsname{\color{black}}%
      \expandafter\def\csname LT6\endcsname{\color{black}}%
      \expandafter\def\csname LT7\endcsname{\color{black}}%
      \expandafter\def\csname LT8\endcsname{\color{black}}%
    \fi
  \fi
  \setlength{\unitlength}{0.0500bp}%
  \begin{picture}(4320.00,4320.00)%
    \gplgaddtomacro\gplbacktext{%
    }%
    \gplgaddtomacro\gplfronttext{%
      \csname LTb\endcsname%
      \put(670,708){\makebox(0,0){\strut{} 1}}%
      \put(982,708){\makebox(0,0){\strut{} 2}}%
      \put(1294,708){\makebox(0,0){\strut{} 3}}%
      \put(1606,708){\makebox(0,0){\strut{} 4}}%
      \put(1918,708){\makebox(0,0){\strut{} 5}}%
      \put(2230,708){\makebox(0,0){\strut{} 6}}%
      \put(2542,708){\makebox(0,0){\strut{} 7}}%
      \put(2854,708){\makebox(0,0){\strut{} 8}}%
      \put(3166,708){\makebox(0,0){\strut{} 9}}%
      \put(3478,708){\makebox(0,0){\strut{}10}}%
      \put(1996,378){\makebox(0,0){\strut{}$\rho\Alt/\rho\Rel$}}%
      \put(342,1128){\makebox(0,0)[r]{\strut{} 1}}%
      \put(342,1397){\makebox(0,0)[r]{\strut{} 2}}%
      \put(342,1666){\makebox(0,0)[r]{\strut{} 3}}%
      \put(342,1935){\makebox(0,0)[r]{\strut{} 4}}%
      \put(342,2203){\makebox(0,0)[r]{\strut{} 5}}%
      \put(342,2471){\makebox(0,0)[r]{\strut{} 6}}%
      \put(342,2740){\makebox(0,0)[r]{\strut{} 7}}%
      \put(342,3009){\makebox(0,0)[r]{\strut{} 8}}%
      \put(342,3277){\makebox(0,0)[r]{\strut{} 9}}%
      \put(342,3546){\makebox(0,0)[r]{\strut{}10}}%
      \put(12,2270){\rotatebox{-270}{\makebox(0,0){\strut{}$\lambda\Rel/\lambda\Alt$}}}%
      \put(3833,993){\makebox(0,0)[l]{\strut{}-6}}%
      \put(3833,1418){\makebox(0,0)[l]{\strut{}-4}}%
      \put(3833,1844){\makebox(0,0)[l]{\strut{}-2}}%
      \put(3833,2269){\makebox(0,0)[l]{\strut{} 0}}%
      \put(3833,2695){\makebox(0,0)[l]{\strut{} 2}}%
      \put(3833,3120){\makebox(0,0)[l]{\strut{} 4}}%
      \put(3833,3546){\makebox(0,0)[l]{\strut{} 6}}%
    }%
    \gplgaddtomacro\gplbacktext{%
    }%
    \gplgaddtomacro\gplfronttext{%
      \csname LTb\endcsname%
      \put(670,708){\makebox(0,0){\strut{} 1}}%
      \put(982,708){\makebox(0,0){\strut{} 2}}%
      \put(1294,708){\makebox(0,0){\strut{} 3}}%
      \put(1606,708){\makebox(0,0){\strut{} 4}}%
      \put(1918,708){\makebox(0,0){\strut{} 5}}%
      \put(2230,708){\makebox(0,0){\strut{} 6}}%
      \put(2542,708){\makebox(0,0){\strut{} 7}}%
      \put(2854,708){\makebox(0,0){\strut{} 8}}%
      \put(3166,708){\makebox(0,0){\strut{} 9}}%
      \put(3478,708){\makebox(0,0){\strut{}10}}%
      \put(1996,378){\makebox(0,0){\strut{}$\rho\Alt/\rho\Rel$}}%
      \put(342,1128){\makebox(0,0)[r]{\strut{} 1}}%
      \put(342,1397){\makebox(0,0)[r]{\strut{} 2}}%
      \put(342,1666){\makebox(0,0)[r]{\strut{} 3}}%
      \put(342,1935){\makebox(0,0)[r]{\strut{} 4}}%
      \put(342,2203){\makebox(0,0)[r]{\strut{} 5}}%
      \put(342,2471){\makebox(0,0)[r]{\strut{} 6}}%
      \put(342,2740){\makebox(0,0)[r]{\strut{} 7}}%
      \put(342,3009){\makebox(0,0)[r]{\strut{} 8}}%
      \put(342,3277){\makebox(0,0)[r]{\strut{} 9}}%
      \put(342,3546){\makebox(0,0)[r]{\strut{}10}}%
      \put(12,2270){\rotatebox{-270}{\makebox(0,0){\strut{}$\lambda\Rel/\lambda\Alt$}}}%
    }%
    \gplbacktext
    \put(0,0){\includegraphics{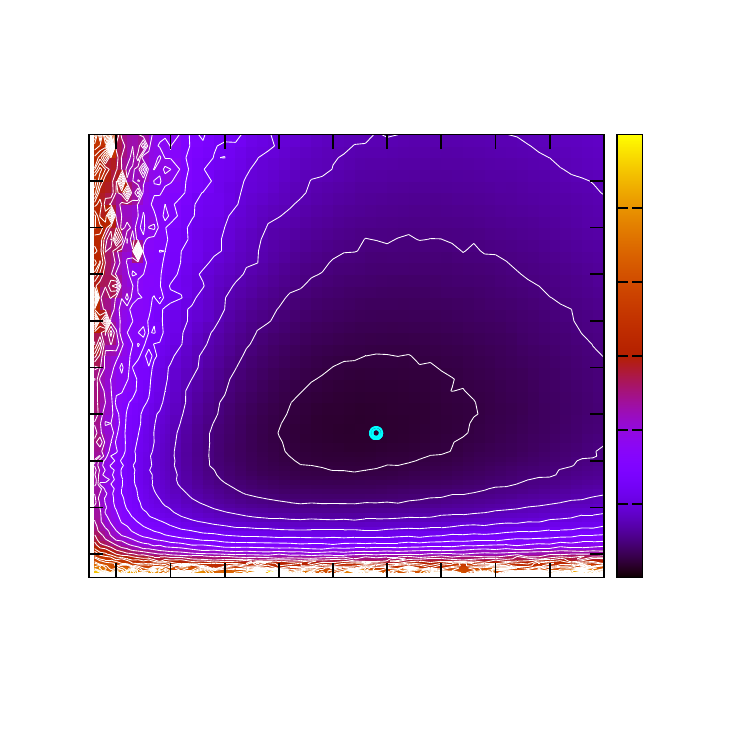}}%
    \gplfronttext
  \end{picture}%
\endgroup
}
\caption[Map of the magnitude of the gain $-\log_2G_\text{num}$ in the $\rho\Alt$--$\frac{1}{\lambda\Alt}$ system]{The magnitude of the gain $-\log_2G_\text{num}$ for the super senior tranche in the $\defPV$ measurement, given in number of Monte Carlo paths as a function of the reweighting parameters: $\rho\Alt$ denotes the altered Poisson process intensity, $\frac{1}{\lambda\Alt}$ the altered expected jump size. The real process uses the real world parameters described in \eqref{par:1} to \eqref{par:3}. The variance decreases for the super senior tranche by increasing the intensity and the jump size to 2-6 times of their original value. \coloronline}
\label{fig:dmap15n}
\end{figure}

\begin{figure}[htbp]
{
\begingroup
  \makeatletter
  \providecommand\color[2][]{%
    \GenericError{(gnuplot) \space\space\space\@spaces}{%
      Package color not loaded in conjunction with
      terminal option `colourtext'%
    }{See the gnuplot documentation for explanation.%
    }{Either use 'blacktext' in gnuplot or load the package
      color.sty in LaTeX.}%
    \renewcommand\color[2][]{}%
  }%
  \providecommand\includegraphics[2][]{%
    \GenericError{(gnuplot) \space\space\space\@spaces}{%
      Package graphicx or graphics not loaded%
    }{See the gnuplot documentation for explanation.%
    }{The gnuplot epslatex terminal needs graphicx.sty or graphics.sty.}%
    \renewcommand\includegraphics[2][]{}%
  }%
  \providecommand\rotatebox[2]{#2}%
  \@ifundefined{ifGPcolor}{%
    \newif\ifGPcolor
    \GPcolortrue
  }{}%
  \@ifundefined{ifGPblacktext}{%
    \newif\ifGPblacktext
    \GPblacktexttrue
  }{}%
  \let\gplgaddtomacro\g@addto@macro
  \gdef\gplbacktext{}%
  \gdef\gplfronttext{}%
  \makeatother
  \ifGPblacktext
    \def\colorrgb#1{}%
    \def\colorgray#1{}%
  \else
    \ifGPcolor
      \def\colorrgb#1{\color[rgb]{#1}}%
      \def\colorgray#1{\color[gray]{#1}}%
      \expandafter\def\csname LTw\endcsname{\color{white}}%
      \expandafter\def\csname LTb\endcsname{\color{black}}%
      \expandafter\def\csname LTa\endcsname{\color{black}}%
      \expandafter\def\csname LT0\endcsname{\color[rgb]{1,0,0}}%
      \expandafter\def\csname LT1\endcsname{\color[rgb]{0,1,0}}%
      \expandafter\def\csname LT2\endcsname{\color[rgb]{0,0,1}}%
      \expandafter\def\csname LT3\endcsname{\color[rgb]{1,0,1}}%
      \expandafter\def\csname LT4\endcsname{\color[rgb]{0,1,1}}%
      \expandafter\def\csname LT5\endcsname{\color[rgb]{1,1,0}}%
      \expandafter\def\csname LT6\endcsname{\color[rgb]{0,0,0}}%
      \expandafter\def\csname LT7\endcsname{\color[rgb]{1,0.3,0}}%
      \expandafter\def\csname LT8\endcsname{\color[rgb]{0.5,0.5,0.5}}%
    \else
      \def\colorrgb#1{\color{black}}%
      \def\colorgray#1{\color[gray]{#1}}%
      \expandafter\def\csname LTw\endcsname{\color{white}}%
      \expandafter\def\csname LTb\endcsname{\color{black}}%
      \expandafter\def\csname LTa\endcsname{\color{black}}%
      \expandafter\def\csname LT0\endcsname{\color{black}}%
      \expandafter\def\csname LT1\endcsname{\color{black}}%
      \expandafter\def\csname LT2\endcsname{\color{black}}%
      \expandafter\def\csname LT3\endcsname{\color{black}}%
      \expandafter\def\csname LT4\endcsname{\color{black}}%
      \expandafter\def\csname LT5\endcsname{\color{black}}%
      \expandafter\def\csname LT6\endcsname{\color{black}}%
      \expandafter\def\csname LT7\endcsname{\color{black}}%
      \expandafter\def\csname LT8\endcsname{\color{black}}%
    \fi
  \fi
  \setlength{\unitlength}{0.0500bp}%
  \begin{picture}(4320.00,4320.00)%
    \gplgaddtomacro\gplbacktext{%
    }%
    \gplgaddtomacro\gplfronttext{%
      \csname LTb\endcsname%
      \put(670,708){\makebox(0,0){\strut{} 1}}%
      \put(982,708){\makebox(0,0){\strut{} 2}}%
      \put(1294,708){\makebox(0,0){\strut{} 3}}%
      \put(1606,708){\makebox(0,0){\strut{} 4}}%
      \put(1918,708){\makebox(0,0){\strut{} 5}}%
      \put(2230,708){\makebox(0,0){\strut{} 6}}%
      \put(2542,708){\makebox(0,0){\strut{} 7}}%
      \put(2854,708){\makebox(0,0){\strut{} 8}}%
      \put(3166,708){\makebox(0,0){\strut{} 9}}%
      \put(3478,708){\makebox(0,0){\strut{}10}}%
      \put(1996,378){\makebox(0,0){\strut{}$\rho\Alt/\rho\Rel$}}%
      \put(342,1128){\makebox(0,0)[r]{\strut{} 1}}%
      \put(342,1397){\makebox(0,0)[r]{\strut{} 2}}%
      \put(342,1666){\makebox(0,0)[r]{\strut{} 3}}%
      \put(342,1935){\makebox(0,0)[r]{\strut{} 4}}%
      \put(342,2203){\makebox(0,0)[r]{\strut{} 5}}%
      \put(342,2471){\makebox(0,0)[r]{\strut{} 6}}%
      \put(342,2740){\makebox(0,0)[r]{\strut{} 7}}%
      \put(342,3009){\makebox(0,0)[r]{\strut{} 8}}%
      \put(342,3277){\makebox(0,0)[r]{\strut{} 9}}%
      \put(342,3546){\makebox(0,0)[r]{\strut{}10}}%
      \put(12,2270){\rotatebox{-270}{\makebox(0,0){\strut{}$\lambda\Rel/\lambda\Alt$}}}%
      \put(3833,993){\makebox(0,0)[l]{\strut{}-4}}%
      \put(3833,1248){\makebox(0,0)[l]{\strut{}-3}}%
      \put(3833,1503){\makebox(0,0)[l]{\strut{}-2}}%
      \put(3833,1758){\makebox(0,0)[l]{\strut{}-1}}%
      \put(3833,2014){\makebox(0,0)[l]{\strut{} 0}}%
      \put(3833,2269){\makebox(0,0)[l]{\strut{} 1}}%
      \put(3833,2524){\makebox(0,0)[l]{\strut{} 2}}%
      \put(3833,2780){\makebox(0,0)[l]{\strut{} 3}}%
      \put(3833,3035){\makebox(0,0)[l]{\strut{} 4}}%
      \put(3833,3290){\makebox(0,0)[l]{\strut{} 5}}%
      \put(3833,3546){\makebox(0,0)[l]{\strut{} 6}}%
    }%
    \gplgaddtomacro\gplbacktext{%
    }%
    \gplgaddtomacro\gplfronttext{%
      \csname LTb\endcsname%
      \put(670,708){\makebox(0,0){\strut{} 1}}%
      \put(982,708){\makebox(0,0){\strut{} 2}}%
      \put(1294,708){\makebox(0,0){\strut{} 3}}%
      \put(1606,708){\makebox(0,0){\strut{} 4}}%
      \put(1918,708){\makebox(0,0){\strut{} 5}}%
      \put(2230,708){\makebox(0,0){\strut{} 6}}%
      \put(2542,708){\makebox(0,0){\strut{} 7}}%
      \put(2854,708){\makebox(0,0){\strut{} 8}}%
      \put(3166,708){\makebox(0,0){\strut{} 9}}%
      \put(3478,708){\makebox(0,0){\strut{}10}}%
      \put(1996,378){\makebox(0,0){\strut{}$\rho\Alt/\rho\Rel$}}%
      \put(342,1128){\makebox(0,0)[r]{\strut{} 1}}%
      \put(342,1397){\makebox(0,0)[r]{\strut{} 2}}%
      \put(342,1666){\makebox(0,0)[r]{\strut{} 3}}%
      \put(342,1935){\makebox(0,0)[r]{\strut{} 4}}%
      \put(342,2203){\makebox(0,0)[r]{\strut{} 5}}%
      \put(342,2471){\makebox(0,0)[r]{\strut{} 6}}%
      \put(342,2740){\makebox(0,0)[r]{\strut{} 7}}%
      \put(342,3009){\makebox(0,0)[r]{\strut{} 8}}%
      \put(342,3277){\makebox(0,0)[r]{\strut{} 9}}%
      \put(342,3546){\makebox(0,0)[r]{\strut{}10}}%
      \put(12,2270){\rotatebox{-270}{\makebox(0,0){\strut{}$\lambda\Rel/\lambda\Alt$}}}%
    }%
    \gplbacktext
    \put(0,0){\includegraphics{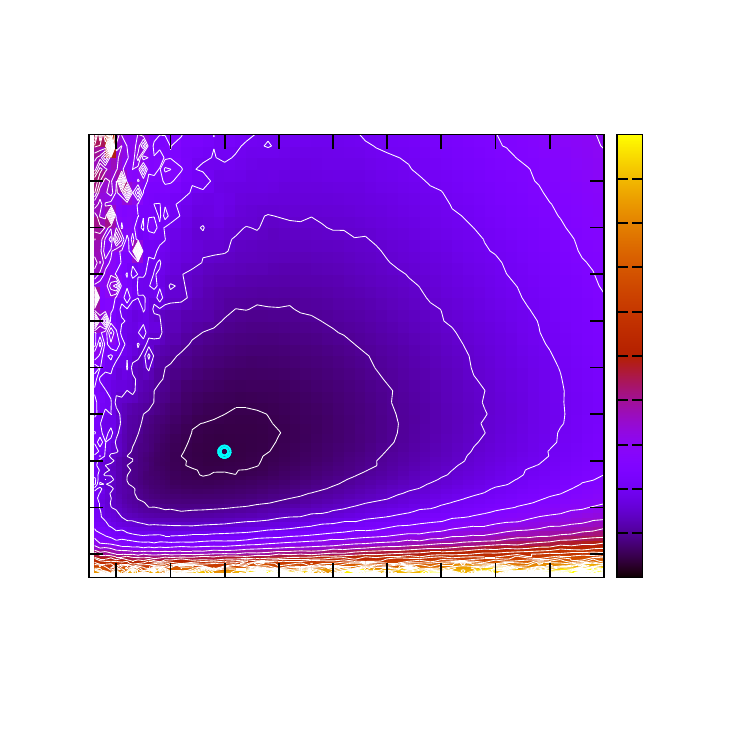}}%
    \gplfronttext
  \end{picture}%
\endgroup
}
\caption[Map of the magnitude of the gain $-\log_2G_\text{time}$ in the
$\rho\Alt$--$\frac{1}{\lambda\Alt}$ system]{The magnitude of the gain
$-\log_2G_\text{time}$ for the super senior tranche in the $\defPV$ measurement, given in computational time
as a function of the reweighting parameters: $\rho\Alt$ denotes the altered
Poisson process intensity, $\frac{1}{\lambda\Alt}$ the altered expected jump
size. The real process uses the real world parameters described in
\eqref{par:1} to \eqref{par:3}. The variance decreases for the super senior
tranche by increasing the intensity and the jump size to 2-4 times of their
original value, but this is less significant than in Figure \ref{fig:dmap15n}. \coloronline}
\label{fig:dmap15p}
\end{figure}

If we look at Figures \ref{fig:dmap15n} and \ref{fig:dmap15p} we can conclude
that the Monte Carlo reweighting method is successful in sparing computational
time. For the pricing of the super senior tranche the total saving in computer
time reaches $91\%$ of the original time without reweighting, which is more than
appealing. (The total saving is defined as $100\%\cdot(1-\frac{1}{G_*})$ where
$*$ stands for number of paths or time.) The saving in the number of paths was
at least $66\%$ for every tranche, worse for the equity tranche. The numerical
results are summarized in Table \ref{tab:sum}.

With some further considerations, one can show that these figures should be
smooth, which is satisfied except for the region of rare but large jumps. This is explained by the relatively small number of paths ($10^5$) being simulated, an order less than for Figures \ref{fig:dlambda15} to \ref{fig:prho15}. This small number of $10^5$ paths is, however, justified, because the full map with this acceptable resolution is calculated in more than an hour on a state-of-the-art computer, the optimal sampling of this map is out of the scope of the current paper.
We would like to assure the reader that the $[0,0.5]$ margins on the maps were
intentionally left out. We see already in the $[0.5,1]$ region the tendency
of increasing variance and the analytical work suggests
that the variance explodes in this model if the altered expected jump size goes
below the half of the original.

In addition to the above mentioned cases, we have analyzed some more extreme, crisis-like situations where $\rho\Rel\approx 1 \frac{\text{event}}{\text{year}}$.
It needed even more computer power because
of the increase in intensity (thus jumps) in a $M=5\text{ year}$ CDO contract. As anticipated, the gain in either number of paths or computational time  is less spectacular, since even the original estimation was not as poor as with the previous, ``calm'' parameter-set.
This simulation showed that the reweighting has its limits, even in the number of paths less than $75\%$ saving was achieved for the super senior tranche and at most $25\%$ or nothing for the others, in contrast to the minimum of $66\%$ of the previous case.

\begin{table*}[tb]
\begin{tabular}{| r r | r r | r r|r | r r|r |}
\hline
\multicolumn{2}{|c|}{Tranche} &
\multicolumn{2}{| c |} {Original values} &
\multicolumn{3}{| c |} {$G_\text{num}$ optimum} &
\multicolumn{3}{| c |} {$G_\text{time}$ optimum} \\
\multicolumn{2}{| c |} {} & \multicolumn{2}{| c |} {in BP rel. to index tr.} &
\multicolumn{2}{| c |} {place} & \multicolumn{1}{| c |} {gain} &
\multicolumn{2}{| c |} {place} & \multicolumn{1}{| c |} {gain} \\
\multicolumn{1}{| c  } {$a$} & \multicolumn{1}{  c |} {$d$} &
\multicolumn{1}{| c  } {$\E(\Xdef)$} & \multicolumn{1}{  c |} {$\sigma(\Xdef)$} &
\multicolumn{1}{| c  } {$\rho\Alt$} & \multicolumn{1}{  c |} {$1/\lambda\Alt$} & \multicolumn{1}{| c |} {value} &
\multicolumn{1}{| c  } {$\rho\Alt$} & \multicolumn{1}{  c |} {$1/\lambda\Alt$} & \multicolumn{1}{| c |} {value} \\
\multicolumn{1}{| c  } {[1]} & \multicolumn{1}{  c |} {[1]} &
\multicolumn{1}{| c  } {[BP]} & \multicolumn{1}{  c |} {[BP]} &
\multicolumn{1}{| c  } {$\big[\frac{\mathrm{event}}{\mathrm{year}}\big]$} &
\multicolumn{1}{  c |} {$\big[\frac{1}{\mathrm{event}}\big]$} & \multicolumn{1}{| c |} {[1]} &
\multicolumn{1}{| c  } {$\big[\frac{\mathrm{event}}{\mathrm{year}}\big]$} &
\multicolumn{1}{  c |} {$\big[\frac{1}{\mathrm{event}}\big]$} & \multicolumn{1}{| c |} {[1]} \\
\hline
0.00 & 0.03 &  58.6 & 115~~ & 0.20 & 0.11~ &  3.0~ & ~0.08(5) & 0.11~ &  1.1~ \\
0.03 & 0.07 &  57.1 & 134~~ & 0.22 & 0.13~ &  3.1~ & ~0.07(5) & 0.13~ &  1.1~ \\
0.07 & 0.10 &  30.8 &  88~~ & 0.23 & 0.16~ &  3.5~ & ~0.07(0) & 0.16~ &  1.2~ \\
0.10 & 0.15 &  35.1 & 121~~ & 0.23 & 0.18~ &  4.3~ & ~0.09(0) & 0.17~ &  1.5~ \\
0.15 & 0.30 &  39.1 & 202~~ & 0.25 & 0.27~ &  9.3~ & ~0.11(0) & 0.25~ &  2.7~ \\
0.30 & 1.00 &   6.8 &  92~~ & 0.28 & 0.38~ & 53.2~ & ~0.16(0) & 0.34~ & 12.4~ \\
0.00 & 1.00 & 227.7 & 606~~ & 0.23 & 0.18~ &  6.0~ & ~0.11(0) & 0.17~ &  1.8~ \\
\hline
\end{tabular}
\vspace{1em}
\caption[The optimum of the gains $G_\text{time}$ and $G_\text{num}$ for $\defPV$ measurement]{The optimum of the gains $G_\text{time}$ and $G_\text{num}$ for $\defPV$ measurement. The reweighting parameters: $\rho\Alt$ denotes the Poisson process intensity, $\frac{1}{\lambda\Alt}$ the expected jump size. The lowest gain appears for the equity tranche ($a=0$, $d=0.03$), the highest for the super senior tranche ($a=0.3$, $d=1$). The real process uses the real world parameters described in \eqref{par:1} to \eqref{par:3} as $\rho\Rel=0.05 \frac{\text{event}}{\text{year}}$, $\frac{1}{\lambda\Rel} = 0.1 \frac{\text{original notional}}{\text{event}}$, $M=5\text{ years}$. The index tranche shows a comportment between the equity and the super senior tranches' comportment.}
\label{tab:sum}
\end{table*}

\section{Conclusions}

In this paper, we have shown the capabilities of importance sampling in estimating the fair price of CDO tranches.
The simple model we covered, enabled us to check our results both analytically and by computer simulation. We showed that this approach is promising in pricing rare events, nevertheless, it has to be treated with care, since even in this basic model, singular behavior emerged.

Future directions include testing the method for more elaborate models (such as \cite{Longstaff2008a} or \cite{Brigo2007a}) and automatizing the optimization procedure, more specifically, gain a parameter set that simultaneously improves all standard tranches.

\section*{Acknowledgement}
We would like to thank Morgan Stanley Hungary for the financial support. We are grateful to
J\'anos Kert\'esz and B\'alint T\'oth for the stimulating discussions. The work
of B.\ V.\ was partially supported by OTKA (Hungarian National Research Fund)
grant K 60708.

\bibliographystyle{apsrev4-1}
%

\end{document}